\documentclass[twocolumn,twocolappendix,trackchanges]{aastex63}

\submitjournal{ApJ}

\shorttitle{CO and H$\alpha$ in NGC 1365}
\shortauthors{Egusa et al.}

\graphicspath{{./}{figures/}}
\usepackage{comment}

\turnoffedit

\begin{document}

\title{CO Excitation and its Connection to Star Formation at 200 pc in NGC 1365}

\author[0000-0002-1639-1515]{Fumi Egusa}
\affiliation{Institute of Astronomy, the University of Tokyo \\
2-21-1 Osawa, Mitaka, Tokyo 181-0015, Japan}
\email{fegusa@ioa.s.u-tokyo.ac.jp}

\author[0000-0002-5973-694X]{Yulong Gao}
\affiliation{School of Astronomy and Space Science, Nanjing University, Nanjing 210093, China}
\affiliation{Institute of Astronomy, the University of Tokyo \\
2-21-1 Osawa, Mitaka, Tokyo 181-0015, Japan}
\affiliation{School of Astronomy and Space Science, University of Science and Technology of China, Hefei 230026, China}

\author[0000-0003-3932-0952]{Kana Morokuma-Matsui}
\affiliation{Institute of Astronomy, the University of Tokyo \\
2-21-1 Osawa, Mitaka, Tokyo 181-0015, Japan}

\author[0000-0003-2390-7927]{Guilin Liu}
\affiliation{School of Astronomy and Space Science, University of Science and Technology of China, Hefei 230026, China}
\affiliation{CAS Key Laboratory for Research in Galaxies and Cosmology, Department of Astronomy, University of Science and Technology of China, Hefei 230026, China}

\author[0000-0002-8868-1255]{Fumiya Maeda}
\affiliation{Institute of Astronomy, the University of Tokyo \\
2-21-1 Osawa, Mitaka, Tokyo 181-0015, Japan}

%% needs to be <250 words
\begin{abstract}
We report high resolution ($2'' \sim 200$ pc) mappings of the central region of the nearby barred spiral galaxy NGC 1365 in the CO(1--0) and CO(2--1) emission lines.
The 2--1/1--0 ratio of integrated intensities shows a large scatter \edit1{(0.15) } with a median value of 
\edit1{0.67. } 
We also calculate the ratio of velocity dispersions and peak temperatures and find that in most cases the velocity dispersion ratio is close to unity and thus the peak temperature ratio is comparable to the integrated intensity ratio.
This result indicates that both CO(1--0) and CO(2--1) lines trace similar components of molecular gas, with their integrated intensity (or peak temperature) ratios reflecting the gas density and/or temperature.
Similar to recent kpc scale studies, these ratios show a positive correlation with a star formation rate indicator (here we use an extinction-corrected H$\alpha$ map), \edit1{suggesting that molecular gas associated with recent star formation is denser and/or warmer.}
\edit1{
We also find that some CO spectra show two peaks owing to complicated kinematics, and such two components likely trace molecular gas at different conditions.
This result demonstrates the importance of spectral fitting to measure integrated intensities and their ratios more accurately.
}
\end{abstract}

%% no longer used
%\keywords{galaxies: individual (NGC 1365) -- }

\section{Introduction} \label{sec:intro}
$^{12}$C$^{16}$O (hereafter CO) is the most abundant molecular species after H$_2$ and thus has been used as a tracer of molecular gas. 
While the lowest rotational transition, CO($J=$1--0), has been regarded as a tracer of bulk cold molecular gas, CO($J=$2--1) has recently become similarly popular because of the high sensitivity of the Atacama Large Millimeter/submillimeter Array (ALMA). 
Currently, 
the ratio (of integrated intensities in K km/s or of peak temperature in K) between the two transitions ($R_{21} =$ 2--1/1--0) is often assumed to be constant \citep[e.g.\ 0.7;][]{SunJ18} for deriving the H$_2$ mass. 
However, brightness of emission lines (and thus their ratios) are dependent of physical conditions.
The line ratios are estimated by different ways.
One example is to assume that the gas is in Local Thermodynamic Equilibrium \citep[LTE, see e.g.][]{Man15}.
Under the optically thick LTE condition, and assuming that both lines come from the same gas component, $R_{21}$ increases with temperature
\edit1{; however, its dynamic range is small -- $R_{21} = 0.64$ for $T = 5.5$ K (corresponding to the upper state energy of CO(1--0)) and approaches to unity for higher temperature.}
If the opacity is lower, 
$R_{21}$ exceeds 1 for temperature higher than 10 K.
However, under the non-LTE condition, the $R_{21}$ dependence is somewhat different. 
Based on calculations with the code provided by \citet{vdTak07}, the ratio becomes high when the H$_2$ volume density and/or kinematic temperature are high.
An expected dynamic range of the ratio is wider than the LTE case and $R_{21}$ can be $\lesssim 0.5$ if density and/or temperature is low.

From high-resolution numerical simulations of GMCs, \citet{Pena18} reported that interstellar radiation field and cosmic ray (CR) ionization rate should have a large and positive impact on observed CO line ratios.
However, cloud averaged ratios calculated by \citet{Bis21} show little dependence on these parameters.
On larger scales, 
\cite{Gong20} found that increasing CR or FUV radiation increases $R_{21}$ from their simulations of galactic disks with kpc size boxes.
At pc scale resolutions, they also showed that the scatter of $R_{21}$ is larger when the density and/or temperature is lower.
Meanwhile, these key parameters are difficult to observe.
\cite{Nara14} utilized their simulations of disk galaxies and galaxy mergers to calculate observed CO line ratios, and fitted them as a function of SFR surface density, which can be more easily measured from observations.
While they found that $R_{21}$ increases with SFR surface density, its dependence is rather flat -- $R_{21}$ increases only by $\sim 0.2$ dex even if SFR surface density increases by 6 dex.
The dynamic range of $R_{21}$ is similar to the optically thick LTE case; however, those from recent observations appear wider (see below) than this prediction.

From an observational point of view, the correlation between $R_{21}$ and gas conditions is still unclear.
In the Milky Way, $R_{21}$ is observed to be higher in the center, spiral arms, and active star forming regions \citep[e.g.,][]{Sawa01,Yoda10,NishiA15}.
While these measurements have been done at high (generally pc-scale or finer) resolutions, most of the studies toward external galaxies are at kpc-scale resolutions.
Previous single point (i.e.\ not spatially-resolved) surveys toward nearby galaxy centers \citep[][]{BC92,BC93a} and xCOLD GASS galaxies \citep[$0.01 < z < 0.05$,][]{Sain17} did not find any strong dependence of the ratio on other physical properties.
However, recent studies on kpc scale variation using large mapping surveys of nearby galaxies have reported mild correlations with galactocentric radius, gas surface density, and SFR \citep[e.g.,][]{Yaji21,denB21,Ler21b}.
These results are qualitatively consistent with the above mentioned theoretical prediction that $R_{21}$ becomes higher when gas is denser and/or warmer.
However, it is noteworthy that ratios outside the optically thick single component LTE prediction, i.e.\ $R_{21} \lesssim 0.5$ or $> 1$, are often observed.
In grand-design spiral galaxies, this variation also appears in an azimuthal direction -- $R_{21}$ tends to be higher at the downstream side of spiral arms \citep{Koda12,Koda20} where star formation is active, although the kpc resolution is marginal to resolve the arm width.
In barred galaxies, the radial dependence is not monotonic because $R_{21}$ is elevated around bar ends, where star formation is active \citep{Mura16,Mae20b,Koda20}.
\citet{DiaG21} did not find a clear correlation between $R_{21}$ and SFR, but it is at least partially due to different beam sizes and thus to a possible overestimation of the intrinsic ratio.
In addition to the variation within a galaxy, $R_{21}$ differs among galaxies. 
From Figure 8 of \citet{Yaji21}, the SFR surface density is one of the important parameters; however, it cannot fully explain the observed $R_{21}$ variation.

Studies on $R_{21}$ at higher resolutions are still limited for external galaxies.
\citet{Sor01} measured $R_{21}$ at a 130 pc resolution for selected positions in the Large Magellanic Cloud (LMC), and found that the ratios within the 30 Dor complex, where SF activity shows a significant variation, are approximately unity.
They interpreted this result as molecular gas before star formation is already dense enough to elevate the ratio.
There are several other studies at similar or higher resolutions \citep{Dru14,Zsc18,Her20}; however, the relationship between $R_{21}$ and SFR has remained unexplored.
An exception is the work by \edit1{\citet{Mae22}}, who derived $R_{21}$ at a 100 pc resolution for the nearby strongly barred galaxy NGC 1300.
While no clear correlation between $R_{21}$ and GMC properties are found, $R_{21}$ positively correlates with H$\alpha$ brightness.
Consistently, $R_{21}$ without H$\alpha$ detection tends to be lower than that with H$\alpha$ detection.
This result appears contrary to the above mentioned work for the 30 Dor complex but is consistent with the kpc scale works. 
Therefore, further studies are needed to understand what controls $R_{21}$.
 
In this paper, we focus on \object{NGC 1365}, which is a nearby \citep[$D = 18.1$ Mpc;][]{Jang18} barred spiral galaxy (SB(s)b) in the southern hemisphere (dec $\sim -36$ deg). 
See \citet{Lind99} for a review of this galaxy.
The presence of a Sy1.8 nucleus \citep{Ver10} and the bar is attributed to the complex gas dynamics around the galactic center \citep[e.g.,][]{Faz19,Gao21}.
CO(1--0) and CO(2--1) observations have been done toward this galaxy \citep[][]{Sandq95,Saka07}, but with a lower resolution, lower sensitivity, and/or smaller area.
The high resolution and sensitivity of ALMA has enabled us to investigate CO distributions and excitation conditions at $2''$ (corresponding to 180 pc at the adopted distance), in the center, bar, and bar-arm transition regions.
CO as well as H$\alpha$ as a star formation tracer data are presented in \S \ref{sec:data}.
\edit1{Masks used to make CO moment maps and their ratio maps are described in \S \ref{sec:mask},}
followed by results based on CO moment maps and comparison with the H$\alpha$ data in \S \ref{sec:result}.
In \S \ref{sec:discussion}, we present results from spectral fitting and discuss the difference from the moment analysis.
\S \ref{sec:summary} gives a summary of this paper.

\section{Data} \label{sec:data}

\subsection{CO(1--0)} \label{sec:data_co10}
The CO(1--0) data are taken from two ALMA projects 
(12m data: 2015.1.01135.S, 
7m and Total Power (TP) data: 2017.1.00129.S).
The Field of Views (FoVs) of both projects are approximately same: about $6.0'\times 3.5'$, covering most of the disk of this galaxy.
The standard data reduction has been performed with CASA \citep{McM07}.

For imaging 12m and 7m data, we exclude the visibility data outside the uv distance range 7--110 k$\lambda$ to make the uv coverages of CO(1--0) and CO(2--1) data similar.
The briggs weighting with robust parameter of 0.5 is selected.
The longest uv distance corresponds to $\simeq 1.9''$, and we set the restoring beam size to $2.0''$.
We utilize the automasking algorithm \citep{Kep20} to limit the area to be cleaned.
The pixel size and channel width are set to $0.2''$ and 5 km/s, respectively.

The TP data are added to the cleaned and primary-beam corrected 12m+7m data via the CASA task {\tt feather}.
Following \citet{Koda20}, we adopt an effective beam size of $56.6''$ for the CO(1--0) TP data.
The missing flux, the fraction of flux undetected by interferometric data, is estimated to be 
\edit1{36\% within the CO(2--1) FoV (described in the next subsection).}
\edit1{
The noise RMS per channel of the combined data is measured along the velocity channel where emission is absent (1320--1370 and 1850--1930 km/s).
Median RMS within the CO(2--1) FoV is 0.19 K.
}
\edit1{For the following analysis, the data cube is binned to $2''$ per pixel.}

\subsection{CO(2--1)}
The CO(2--1) data are taken from the ALMA project: 2013.1.01161.S.
The FoV is about $2'\times 3'$, not covering the entire disk but the center, bar, and bar-arm transition of this galaxy.
Data reduction and imaging are same with CO(1--0), except (1) rescaling the clean residual map and (2) convolving the TP data. 

Although the selected uv distance range is same for both the lines, the data density in uv plane is slightly different.
To be more specific, short baseline data (mainly from 7m) are less populated compared to the CO(1--0) data.
This results in a difference between dirty and clean beam areas.
As described in \citet{JvM95} and \citet{Koda19}, this difference causes over- or under-estimation of the clean residuals.
\edit1{
Following Eq.\ (22) of \citet{Koda19}, ratios of clean and dirty beam areas from the dirty, clean, and residual data cubes are calculated, and a typical ratio of approximately 3 is found where emission is clearly detected.
We thus rescale the CO(2--1) clean residual by a factor of 3.
}
Note that this only affects the estimate of noise RMS with no change in fluxes of cleaned components.

The effective beam size of the CO(2--1) TP data is adopted to be $28.3''$.
The CO(2--1) TP data are then convolved to match the TP beam size of the CO(1--0) data \citep[c.f.][]{Koda20} before being added to the 12m+7m data via {\tt feather}.
The missing flux within the CO(2--1) FoV is 31\%.
\edit1{As for CO(1--0), the noise RMS per channel of the combined data is measured along the velocity, and its median value is 0.13 K.}
\edit1{In addition, the data cube is binned to $2''$ per pixel.}

\subsection{H$\alpha$}
The PHANGS-MUSE project provides fully calibrated data cubes and maps of 19 nearby galaxies including NGC 1365 \citep{Ems21}.
From their data archive\footnote{\url{https://www.canfar.net/storage/vault/list/phangs/RELEASES/PHANGS-MUSE/DR1.0}}, we retrieved H$\alpha$ and H$\beta$ maps at $1.15''$ resolution.
Note that [N II] lines around H$\alpha$ are fitted separately (i.e., no correction for [N II] contamination in H$\alpha$ is necessary) and that these maps are already corrected for the MW foreground contribution.
The observed flux ratio of H$\alpha$ to H$\beta$ is used to correct for internal extinction to the H$\alpha$ emission.
We adopted parameters in Table 2 of \citet{Cal01} for this calculation, and excluded pixels where S/N $< 4$ for either of the two emissions.
If calculated extinction becomes negative, no correction is applied.
Median values of $A_{\rm H\alpha}$ and of the S/N of corrected H$\alpha$ flux are 0.4 mag and 11, respectively.
While \citet{Gao21} corrected for the AGN contribution to H$\alpha$ brightness, this correction is not applied in this study because its fraction measured by \citet{Gao21} appears low ($\lesssim 30$\%) and relatively uniform across where CO(1--0) is detected.

\edit1{For the following analysis, the extinction corrected H$\alpha$ map is smoothed and then regridded to make the angular resolution and pixel size the same as those of the CO datasets.}

\begin{figure*}
\includegraphics[scale=0.5,trim=0 0 10 0,clip]{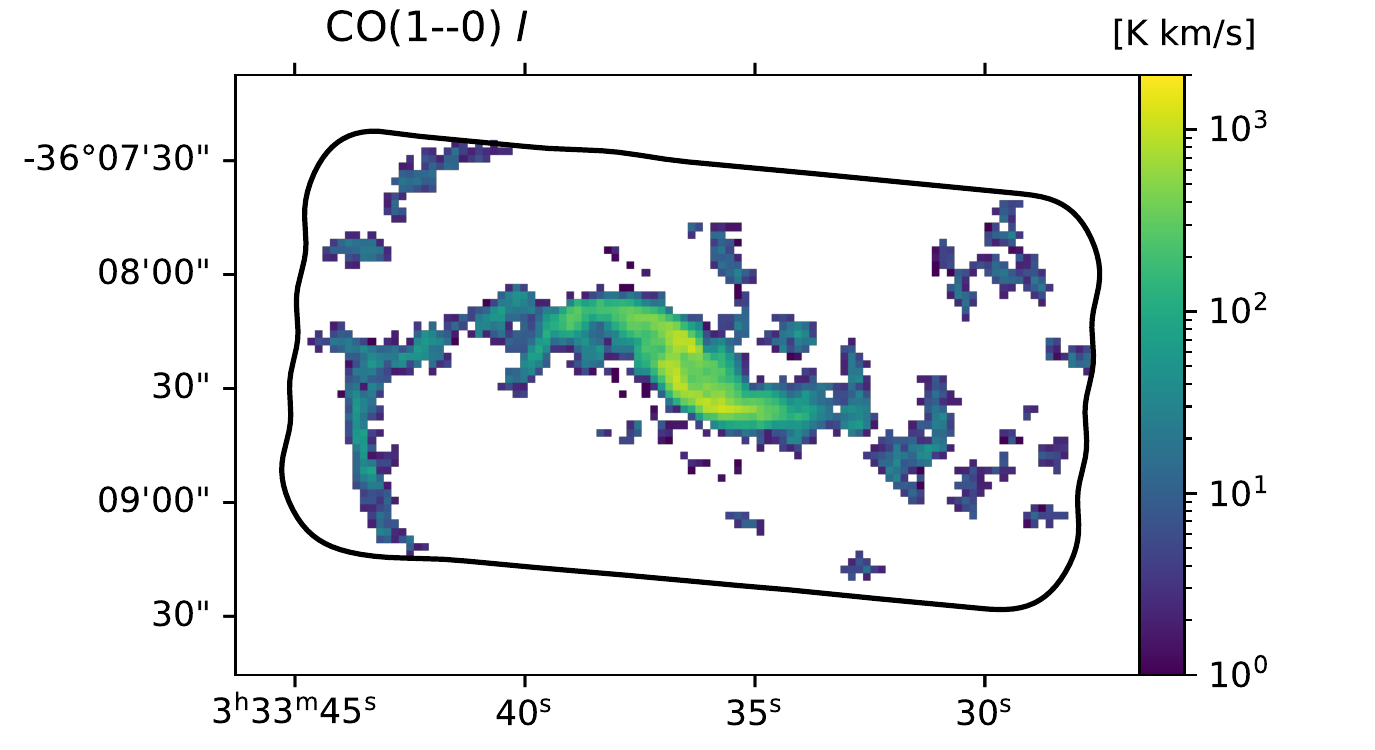}
\includegraphics[scale=0.5,trim=60 0 10 0,clip]{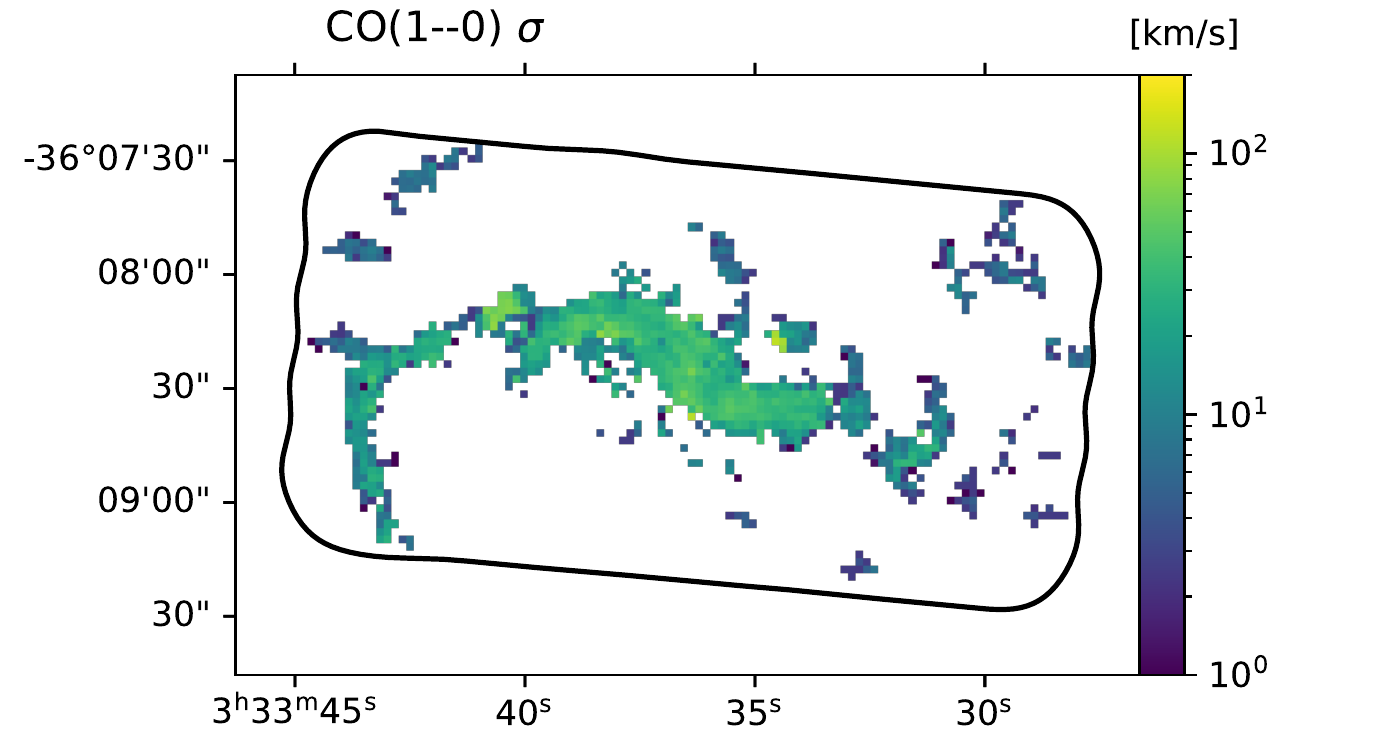}
\includegraphics[scale=0.5,trim=61 0 0 0,clip]{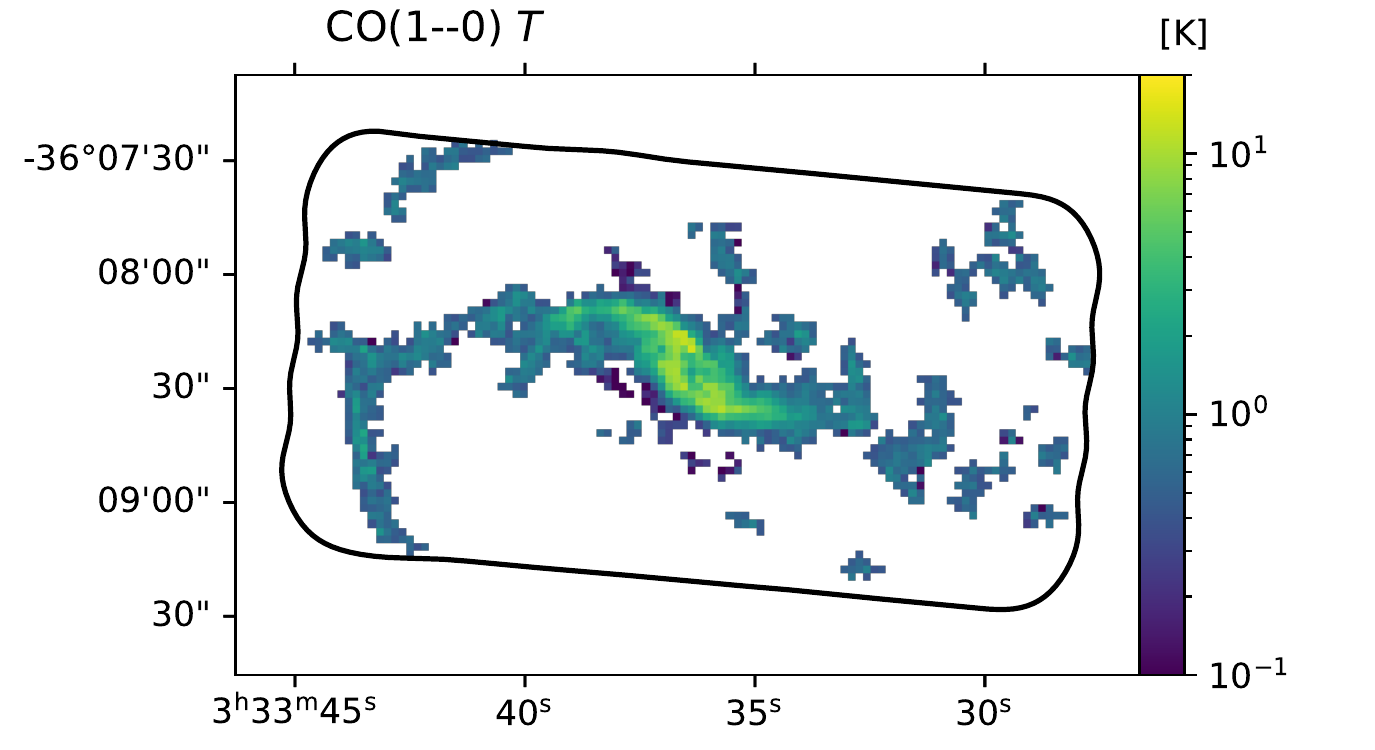}\\
\includegraphics[scale=0.5,trim=0 0 10 0,clip]{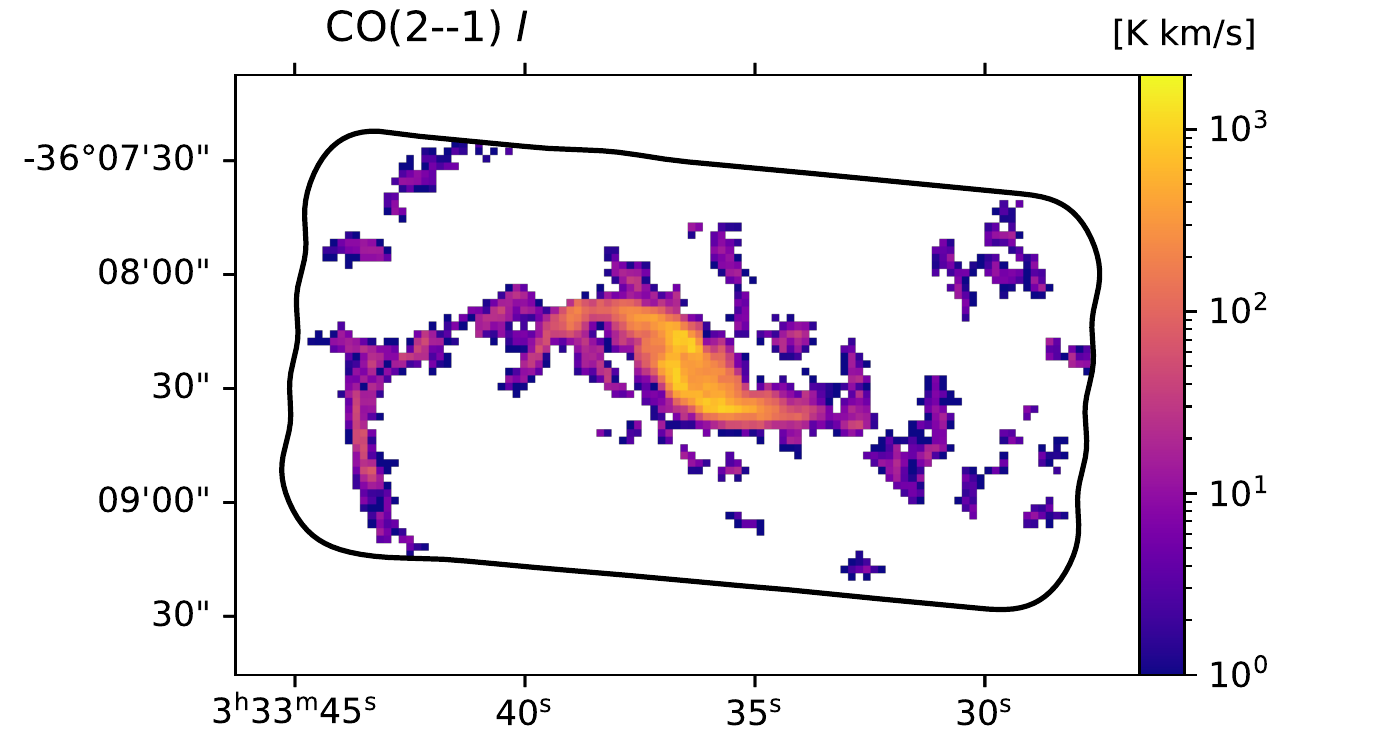}
\includegraphics[scale=0.5,trim=60 0 10 0,clip]{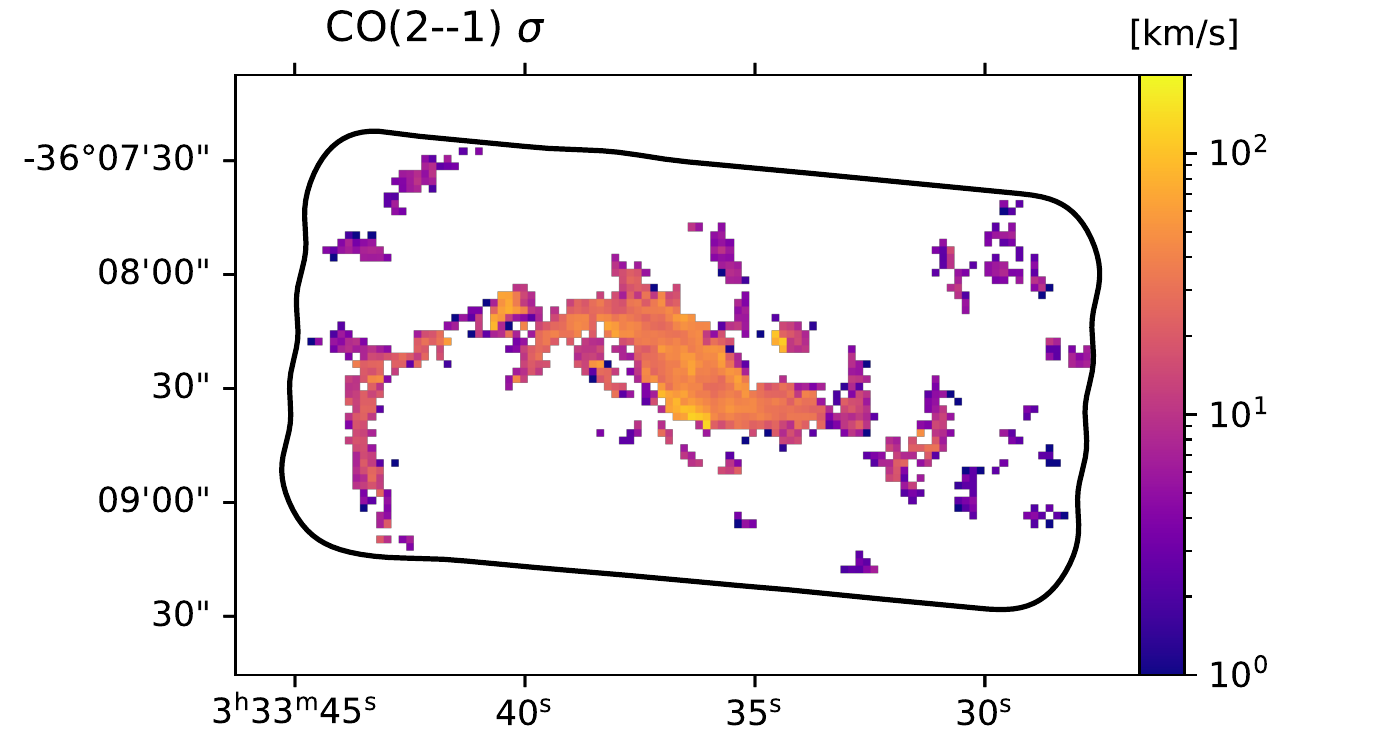}
\includegraphics[scale=0.5,trim=61 0 10 0,clip]{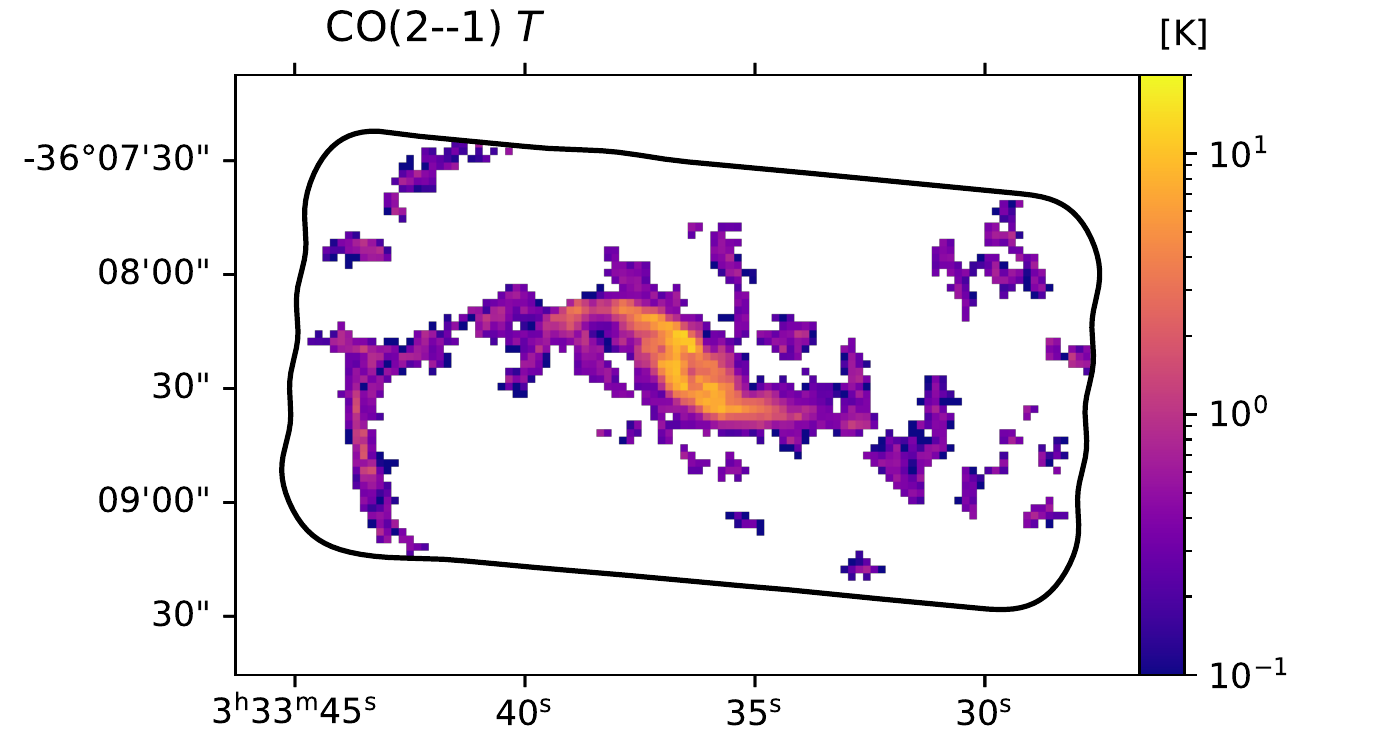}\\
\includegraphics[scale=0.5,trim=0 0 0 0,clip]{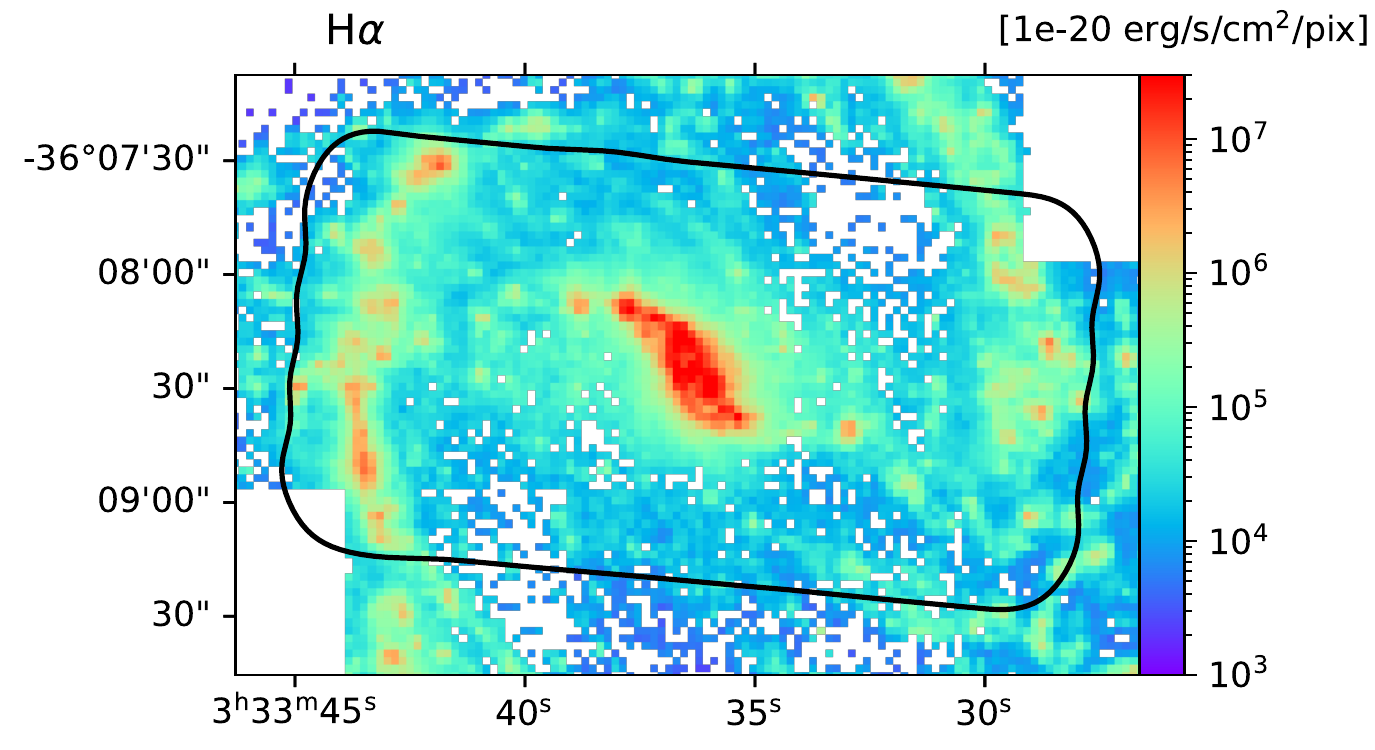}
\includegraphics[scale=0.5,trim=60 0 10 0,clip]{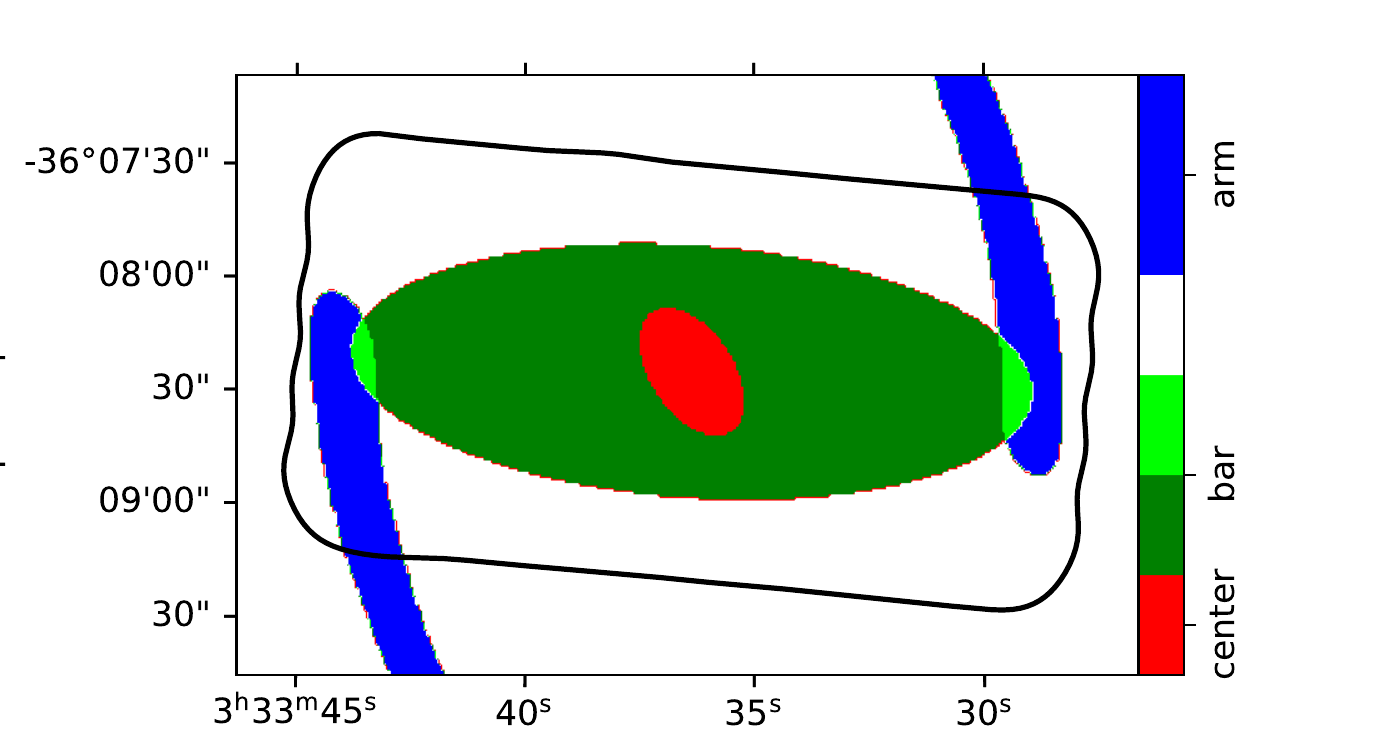}
\caption{
\edit1{
Top row: integrated intensity ($I$), velocity dispersion ($\sigma$), and peak temperature ($T$) of CO(1--0).
Second row: same as the top row, but of CO(2--1).
Third row: extinction corrected brightness of H$\alpha$, and environmental masks from \citet{Que21}.}
In all panels, the black solid curve represents the CO(2--1) FoV.
}
\label{fig:show_images}
\end{figure*}

\edit1{
\section{Masks for calculating moments and ratios}\label{sec:mask}
In general, when creating moment maps from 3D data cubes, excluding noisy voxels is important to improve the S/N of output maps.
For masking these voxels, we adopt a strategy first presented by \citet{Roso06} for cloud identification, which was recently used for making moment maps \citep[e.g.,][]{Mae20b} -- (i) we identify voxels with S/N $> 4$ in at least two adjacent velocity channels in the input cube, (ii) extend the area until S/N becomes $< 2$ (i.e., include all adjacent voxels with S/N $\ge 2$).
This process is performed for each CO data cubes and thus two 3D cube masks are generated.
The two masks are then combined by logical OR, i.e.\ voxels included in either of the two masks are included in the final 3D cube mask.
We use this final 3D cube mask to make moment maps: integrated intensity (moment 0, $I$), intensity-weighted velocity dispersion (moment 2, $\sigma$), and peak temperature (called ``moment 8'' in CASA, $T$).
These moment maps for CO(1--0) and CO(2--1) are presented in Figure \ref{fig:show_images}, together with the smoothed and regridded image of extinction corrected H$\alpha$.
Using the same 3D cube mask and thus the same velocity ranges for both CO lines is important to calculate their ratios.
}

\begin{figure*}
\includegraphics[trim=0 0 0 0,clip,scale=0.5]{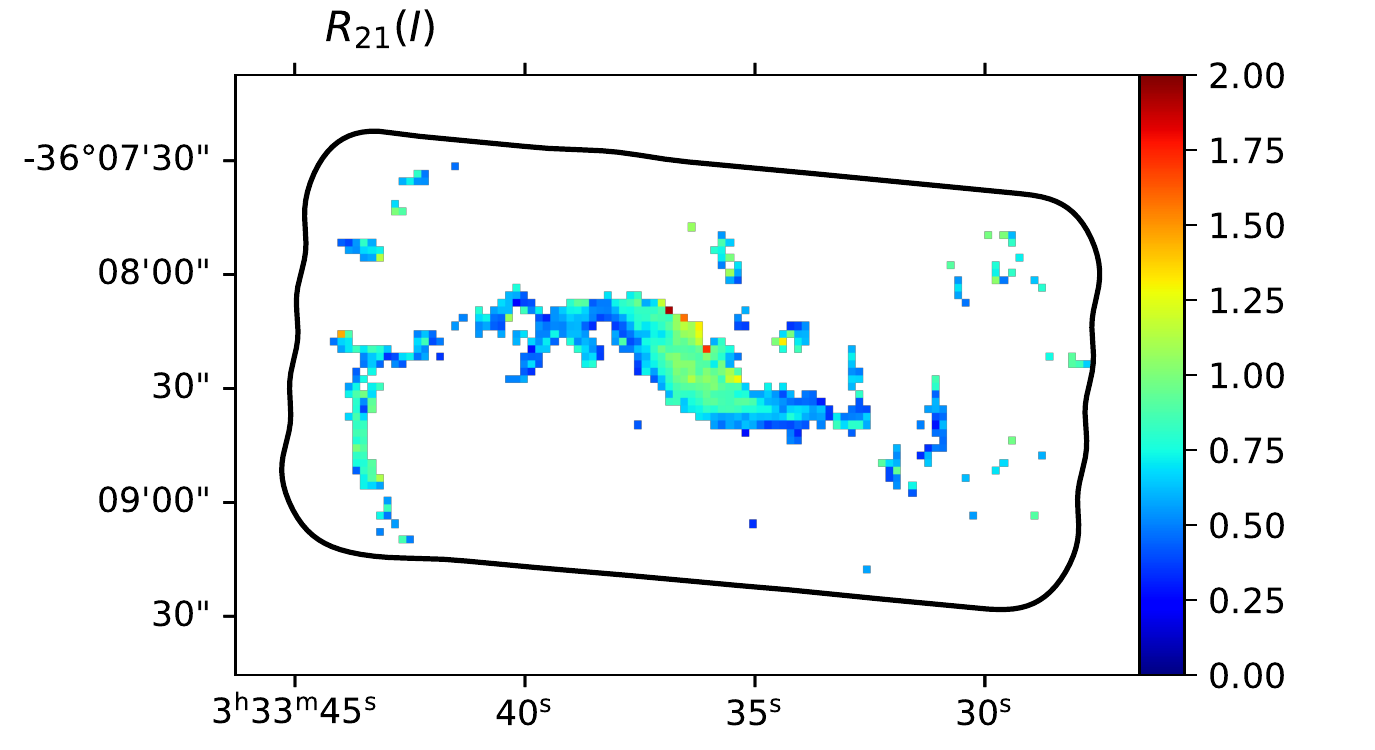}
\includegraphics[trim=61 0 0 0,clip,scale=0.5]{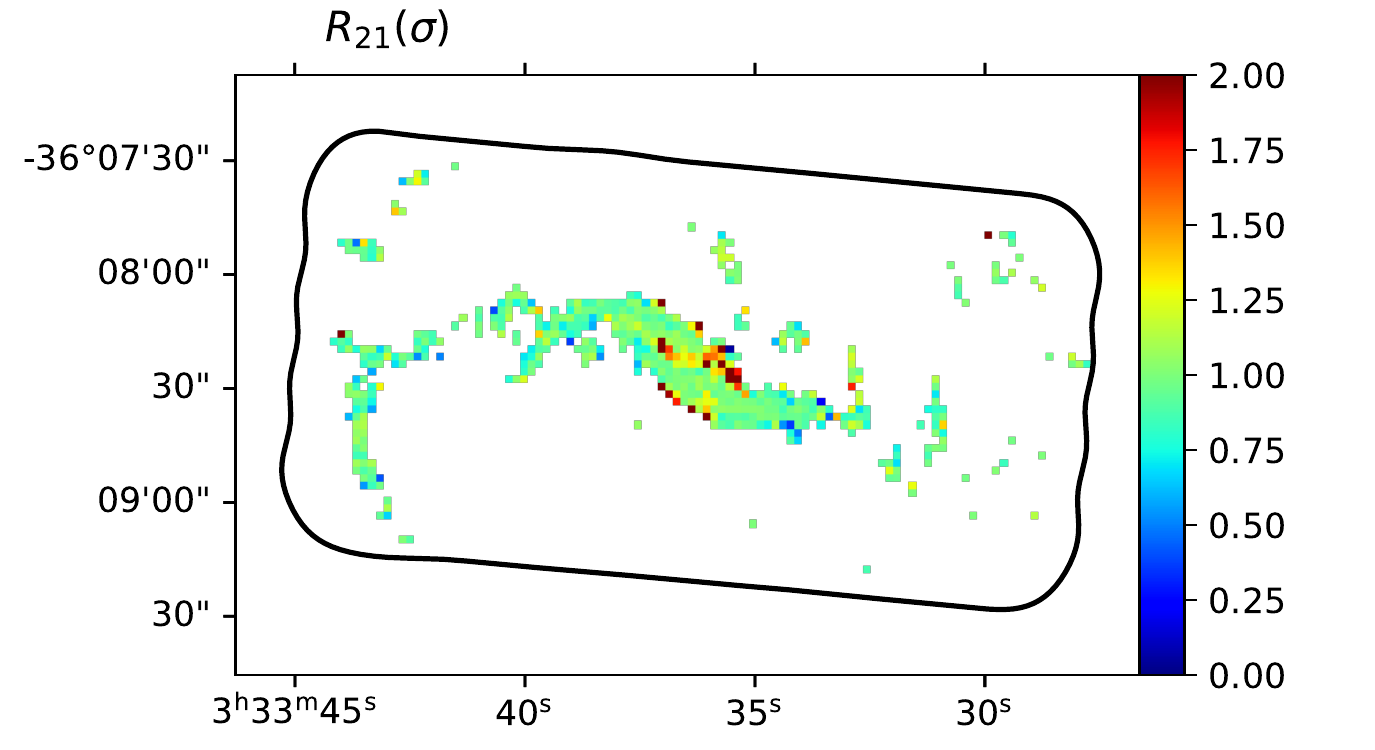}
\includegraphics[trim=61 0 0 0,clip,scale=0.5]{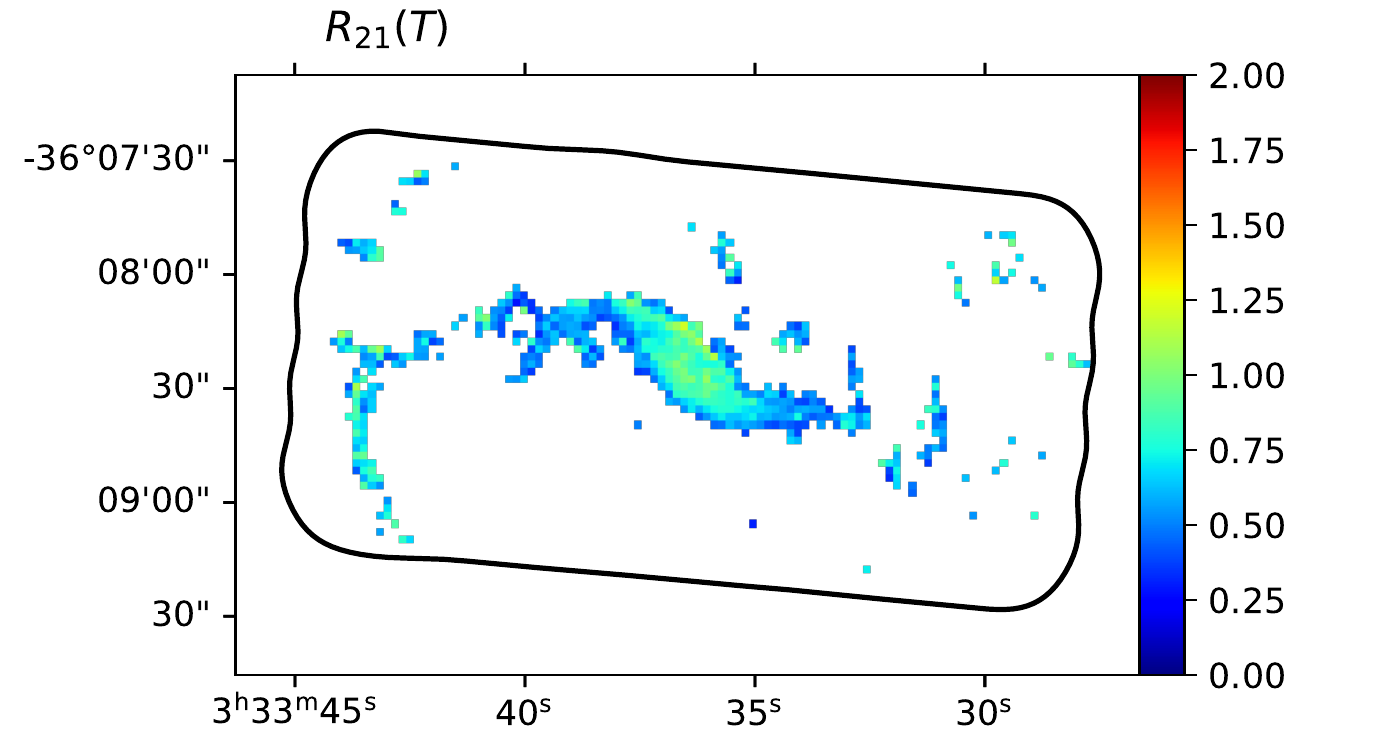}\\
\includegraphics[trim=0 0 0 0,clip,scale=0.5]{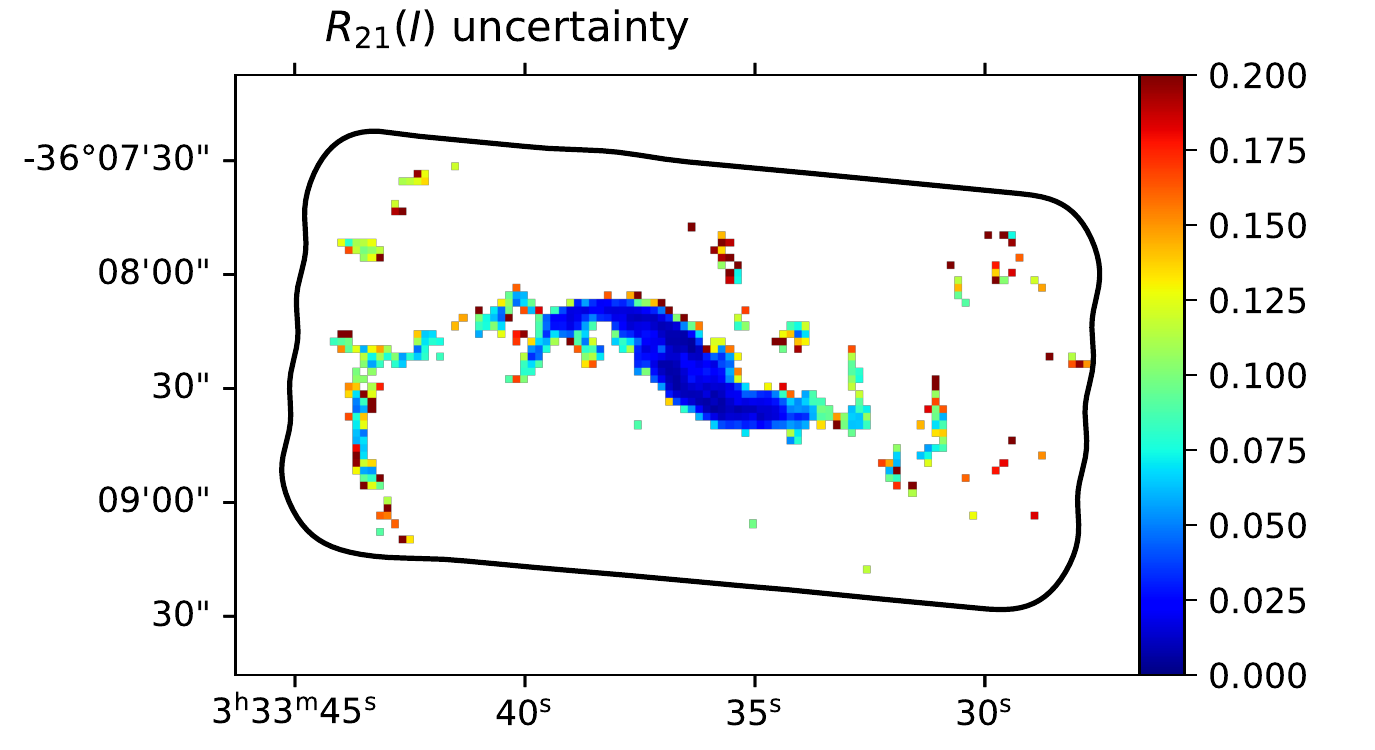}
\includegraphics[trim=61 0 0 0,clip,scale=0.5]{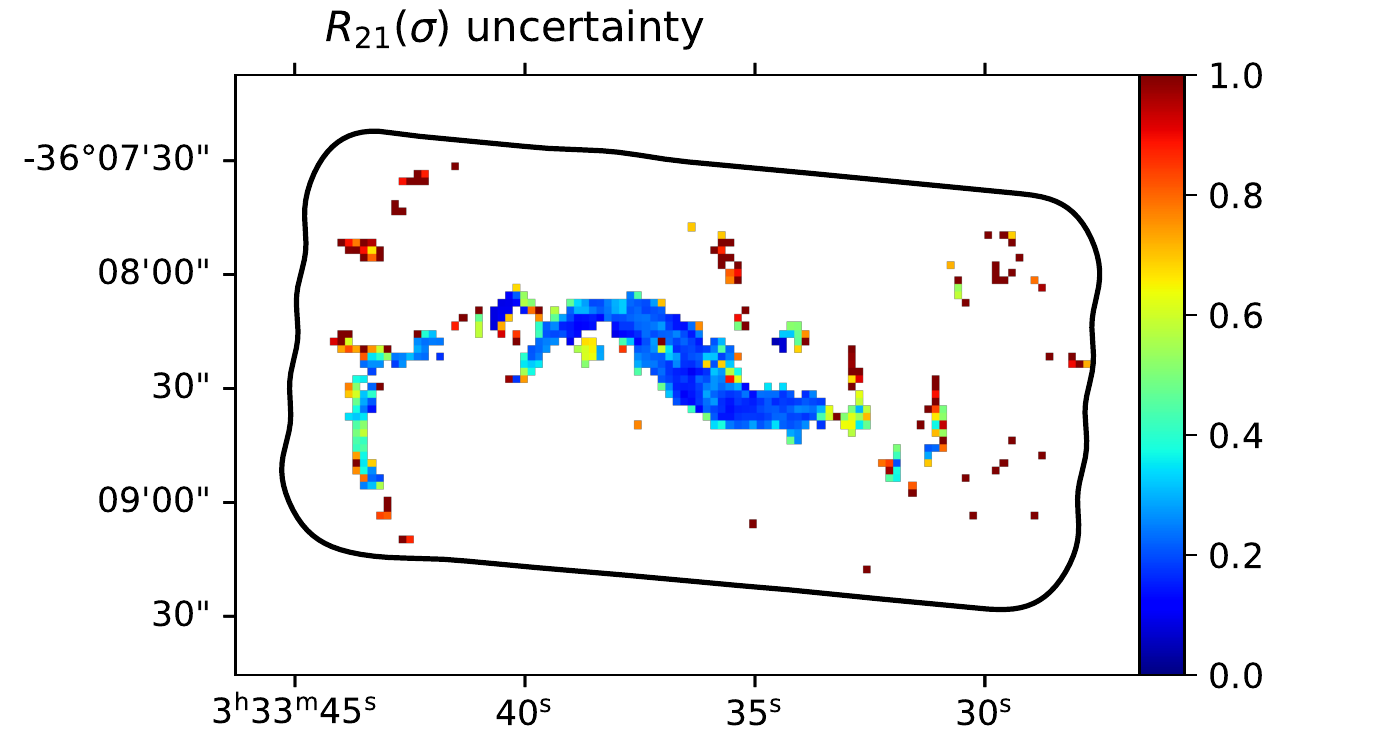}
\includegraphics[trim=61 0 0 0,clip,scale=0.5]{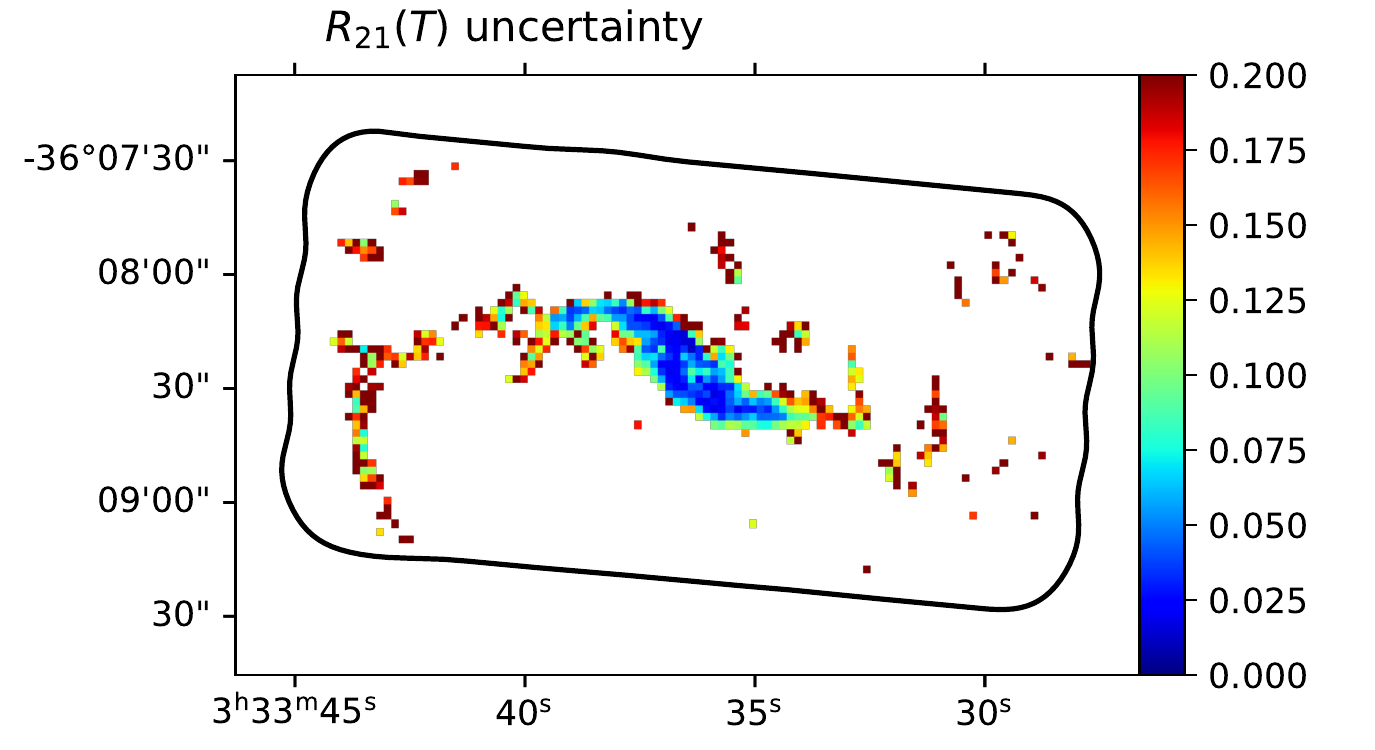}
\caption{ 
\edit1{
Top row: $R_{21}(I)$, $R_{21}(\sigma)$, and $R_{21}(T)$, from left to right.
Bottom row: same as the top row, but for their uncertainties.
}
Note that color ranges are different 
\edit1{among the uncertainties.}
In all panels, the black solid curve represents the CO(2--1) FoV.
}
\label{fig:show_images_R21}
\end{figure*}

\edit1{
For calculating a ratio of these moment maps, another mask is needed to exclude low S/N pixels in moment maps.
We first create a noise RMS map for each $I$ map from the noise RMS per channel and the number of channels inside the final 3D cube mask (i.e.\ used for moment calculation), 
and then make an S/N map for each $I$.
The final 2D mask for ratio calculation is defined by including pixels where both of S/N($I$) satisfy $> 4$.
The 2--1/1--0 ratio of $I$, $\sigma$, and $T$ ($R_{21}(I)$, $R_{21}(\sigma)$, and $R_{21}(T)$, respectively) calculated with this final 2D mask are presented in Figure \ref{fig:show_images_R21}.
The effect of the additional 2D mask can be seen as more blank (i.e., masked) pixels in the ratio maps (Figure \ref{fig:show_images_R21}) compared to that in the moment maps (Figure \ref{fig:show_images}).
}

\edit1{
Uncertainties in these ratios, presented in the bottom row of Figure \ref{fig:show_images_R21}, are calculated by propagating uncertainties in the corresponding moment maps.
For $R_{21}(I)$ uncertainties, the noise RMS maps of $I$ that are used to make the final 2D mask are used.
For $R_{21}(T)$ uncertainties, the maps of noise RMS per channel described in \S \ref{sec:data_co10} are used.
For $R_{21}(\sigma)$ uncertainties, estimating uncertainties in $\sigma$ is not straightforward.
So, we conservatively adopt the channel width (i.e., 5 km/s) as a constant uncertainty in $\sigma$ for both lines.
}

\section{Results}\label{sec:result}
As seen in the maps of CO(1--0) and CO(2--1) in Figure \ref{fig:show_images}, molecular gas distribution and condition traced by these two transitions are rather similar.
\edit1{
In this figure, the environmental masks defined in \citet{Que21} are also presented to delineate the center, bar, and spiral arm regions, while the eastern arm observed in CO and H$\alpha$ appears more extended to the north than this definition.
}
Molecular gas is detected in the bar and spiral arms.
In the leading edges of the bar, both velocity dispersion ($\sigma$) and peak temperature ($T$) are large, so that the integrated intensity ($I$) and thus the gas column density (which is generally calculated from $I$) are also large.
The H$\alpha$ map\edit1{, which covers most of the CO(2--1) FoV,} shows a similar distribution; however, spiral arms appear wider and are easier to trace compared to those in CO maps (at least partially due to a better sensitivity of the H$\alpha$ data).

\subsection{CO(1--0) and CO(2--1) (or $R_{21}$)}\label{sec:r21}
\begin{figure*}[ht!]
\plotone{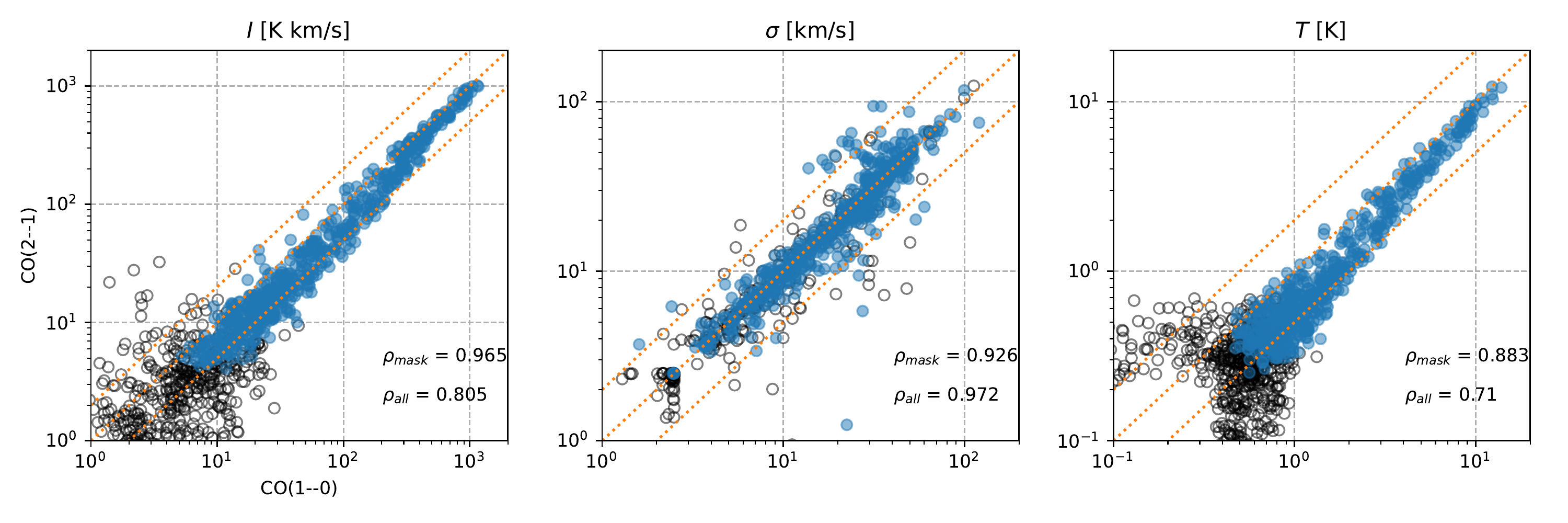}
\caption{(Left) $I$ of CO(2--1) against that of CO(1--0). 
Each data point corresponds to a binned pixel with a size of $2''$.
\edit1{
Gray open circles correspond to pixels excluded by the image mask for ratio calculation.
Orange dotted lines indicate constant ratios of $R_{21} =$ 2, 1, and 0.5.
In the bottom right corner, $\rho_{\rm mask}$, which is Spearman's correlation coefficient with blue data points (i.e., inside the mask), and $\rho_{\rm all}$, which is with all data points (including those outside the plot area), are shown.
}
(Middle) same as the left panel, but for $\sigma$.
(Right) same as the left panel, but for $T$.
\label{fig:plot_moms}}
\end{figure*}

In Figure \ref{fig:plot_moms}, correlations between CO(2--1) and CO(1--0) for $I$, $\sigma$, and $T$ are presented.
\edit1{
Blue filled and gray open circles correspond to pixels included and excluded by the final 2D mask for calculating ratios, respectively.
Spearman's correlation coefficients for blue and all data points (including those outside the plot area), $\rho_{\rm mask}$ and $\rho_{\rm all}$, respectively, are shown in the bottom left corner of each panel.
For $I$ and $T$, $\rho_{\rm mask}$ is larger than $\rho_{\rm all}$, which indicates that the 2D mask can indeed remove low S/N pixels showing a larger scatter.
Orange dotted lines indicate constant ratios (i.e., $R_{21}$) of 2, 1, and 0.5.
}
In general, correlations are tight and the ratios are close to unity, which is consistent with the above statement from CO maps
-- molecular gas distribution and condition traced by CO(1--0) and CO(2--1) are rather similar.

\edit1{
The spatial distribution of $R_{21}(I)$ in Figure \ref{fig:show_images_R21} shows a significant variation within the FoV.
Its median and median absolute deviation (MAD) as a measure of its scatter are 0.67 and 0.15, respectively. 
Higher ratios are found around the center and spiral arms, while lower ratios are found in the outer region of the bar.
}
This non-monotonic behavior is similar to that found in other barred galaxies \citep[e.g.,][]{Mura16,Mae20b,Koda20}.
\edit1{
Meanwhile, the radial variation of $R_{21}(I)$ uncertainties is more systematic with lower values toward the center, and median of 0.07.
Behavior of $R_{21}(T)$ is qualitatively the same as $R_{21}(I)$, and its median, MAD, and median uncertainty are 0.63, 0.13, and 0.14, respectively.
We
}
should note that our observations are biased to brighter components and thus to higher ratios.
\citet{Pena17} stated that lower $R_{21}$ comes from a faint area of a GMC, which may not be detected due to the CO sensitivity limit.
To discuss azimuthal variation, deeper CO observations are necessary.
On the other hand, no clear radial trend of $R_{21}(\sigma)$ is found
\edit1{
with its median, MAD, and median uncertainty being 0.98, 0.09, and 0.30, respectively.
The smallest MAD reflects its uniformity within the FoV; however, extremely high values ($R_{21}(\sigma) > 2$) are preferentially found around the center.
}

\begin{figure*}[ht!]
\plotone{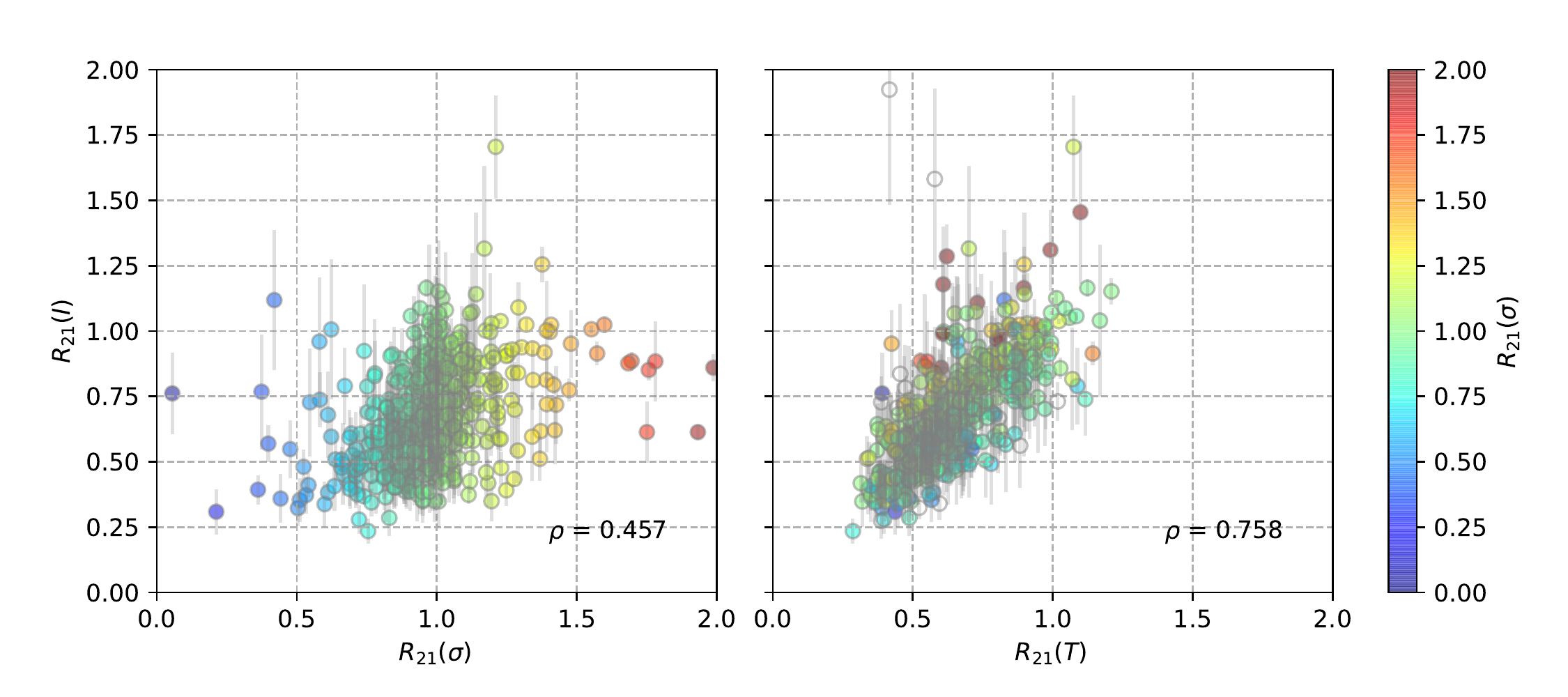}
\caption{Correlation between the three $R_{21}$ values. $R_{21}(I)$ is plotted against $R_{21}(\sigma)$ (left) and $R_{21}(T)$ (right).
\edit1{
Symbols are color-coded by $R_{21}(\sigma)$ values.
Open circles in the right panel correspond to pixels where $R_{21}(\sigma)$ measurements failed (and thus do not appear in the left panel).
}
\label{fig:plot_ratios}}
\end{figure*}

In Figure \ref{fig:plot_ratios}, we compare $R_{21}(I)$, $R_{21}(\sigma)$, and $R_{21}(T)$.
For the clarity of plots, error bars of $R_{21}(\sigma)$ and $R_{21}(T)$ are omitted.
This figure shows majority of data points have $R_{21}(\sigma)$ close to unity, i.e.\ their line widths are similar for both transitions.
For these points, $R_{21}(I)$ is 
\edit1{0.25}--1.0 and determined by $R_{21}(T)$.
\edit1{
The large correlation coefficient ($\rho \simeq 0.8$) reflects this strong correlation between $R_{21}(I)$ and $R_{21}(T)$.
It is noteworthy that a certain portion of this correlation go below the lower limit of the optically-thick LTE case ($\simeq 0.5$) and that some of such data points show $R_{21}(\sigma) < 1$.
This result suggests that molecular gas is not in the LTE condition and/or that the two CO lines are emitted from different parts of molecular gas.
One possibility is that CO(1--0) traces more extended (less dense and cold) part of a molecular cloud while CO(2--1) traces its inner part.
In this figure, symbols are color-coded by $R_{21}(\sigma)$ values.
As naturally expected, data points in redder color ($R_{21}(\sigma) \gtrsim 1.5$) tend to reside in the $R_{21}(I) > R_{21}(T)$ area in the right panel.
These data points also suggest the possible difference of CO emitting regions, but in the other way.
For example, a CO(2--1) line profile can be wider if there is an external heating source.
However, their error bars are relatively large and thus need to be further investigated.
In addition, some outliers are marked by non-colored open circles, which indicate that $R_{21}(\sigma)$ measurements have failed.
This is due to a failure in calculating $\sigma$ values for either of the two CO lines, where spectra show multiple-peak profiles, negative features, and/or faint peaks.
We have confirmed that the $R_{21}(I)$ uncertainties (bottom left of Figure \ref{fig:show_images_R21}) calculated in these pixels appropriately reflect their spectra qualities and, therefore, have decided not to exclude them from the following analysis.
As the fraction of such pixels is as small as 5\%, our results are independent of this decision.
}

\subsection{$R_{21}$ vs H$\alpha$}\label{sec:r21_Ha}
\begin{figure}[ht!]
\includegraphics[trim=0 20 0 0,clip,width=\linewidth]{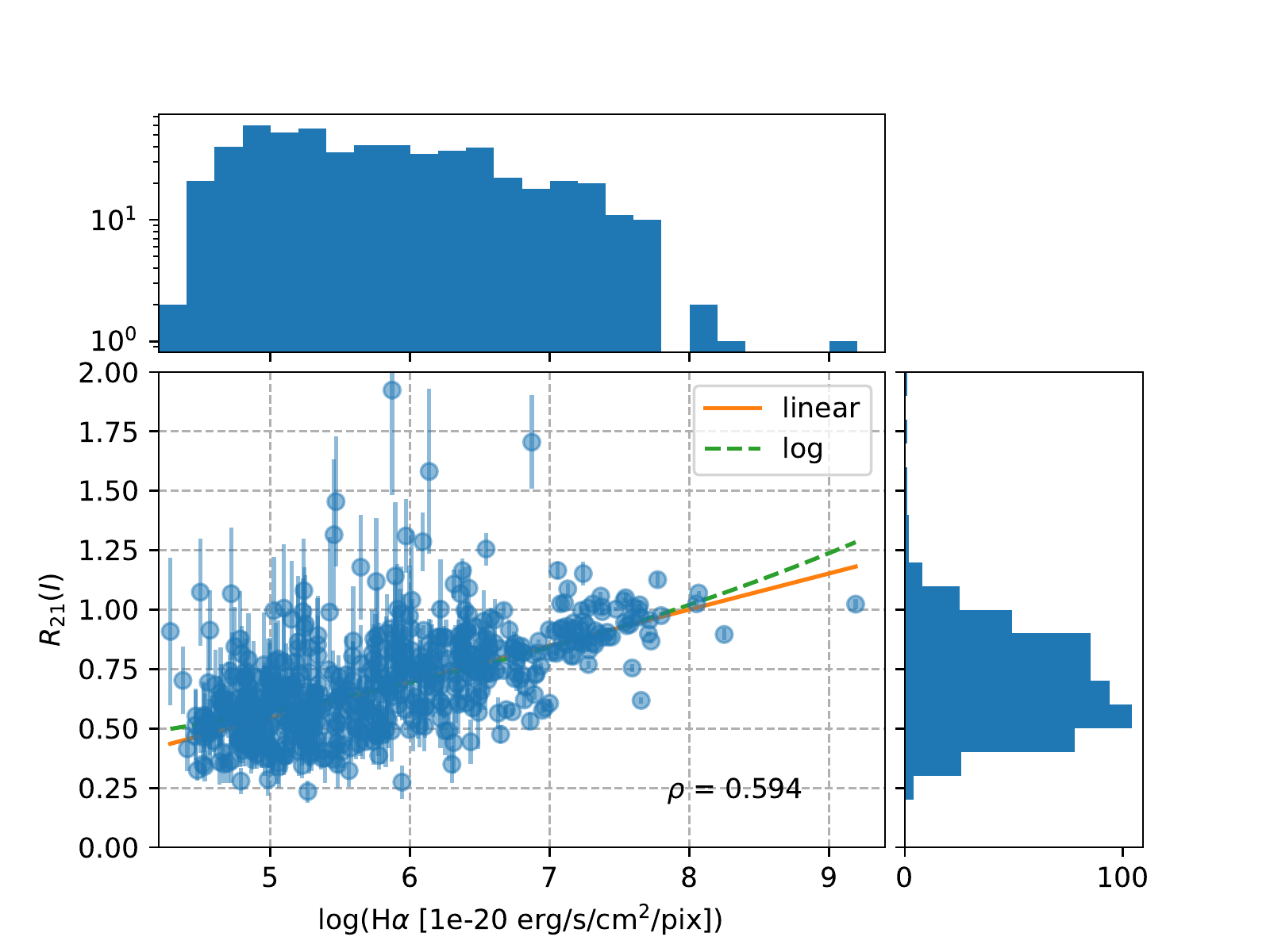}
\includegraphics[trim=0 20 0 131,clip,width=\linewidth]{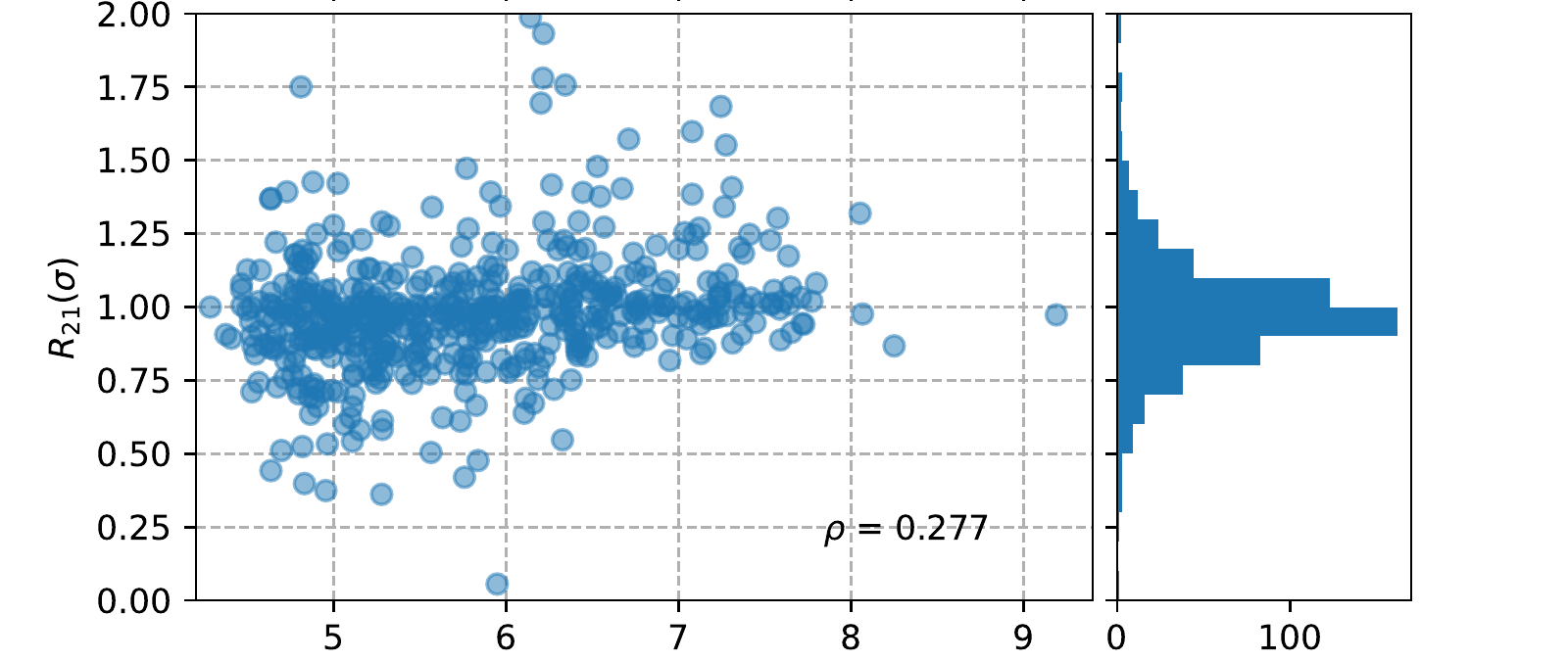}
\includegraphics[trim=0 0 0 131,clip,width=\linewidth]{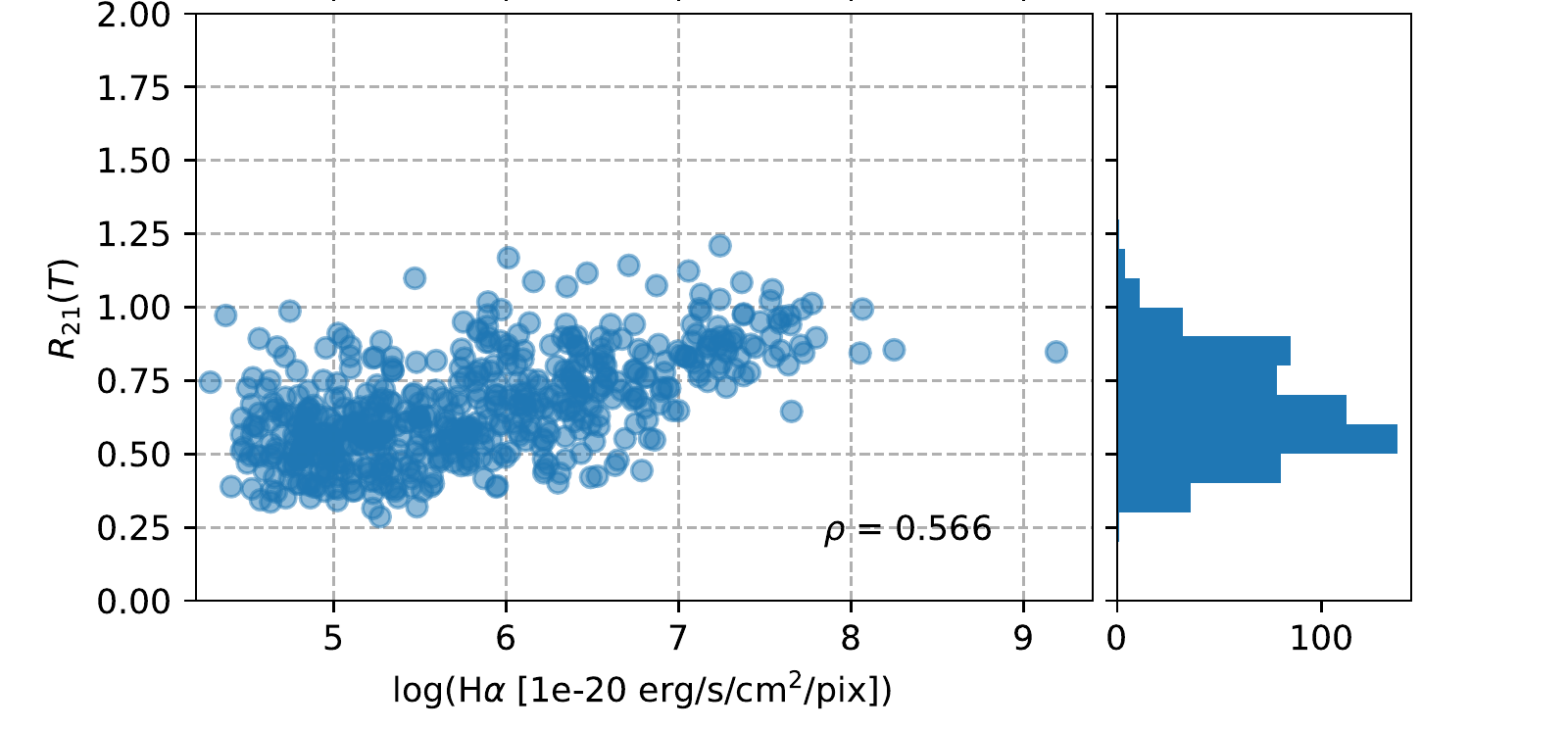}
\caption{$R_{21}(I)$ (top), $R_{21}(\sigma)$ (middle), and $R_{21}(T)$ (bottom) against log(H$\alpha$). 
Additional panels on top and right present histograms of H$\alpha$ and corresponding $R_{21}$ values, respectively.
\edit1{The orange solid and green dashed lines in the top plot show the two fitting results described in \S \ref{sec:r21_Ha}.}
\label{fig:R21_SFR}}
\end{figure}

In Figure \ref{fig:R21_SFR}, three $R_{21}$ values are plotted against the extinction-corrected H$\alpha$ flux (hereafter we just refer it to as ``H$\alpha$ flux'').
Histograms of these values are added to corresponding axes.
Only pixels where both $R_{21}$ and H$\alpha$ are measured are included in these plots.
As for Figure \ref{fig:plot_ratios}, error bars of $R_{21}(\sigma)$ and $R_{21}(T)$ are not plotted.
From this figure, $R_{21}(I)$ and $R_{21}(T)$ are clearly positively correlated with H$\alpha$ flux \edit1{($\rho \simeq 0.6$)}, while $R_{21}(\sigma)$ does not depend on H$\alpha$ \edit1{($\rho \simeq 0.3$)}.
While the H$\alpha$ flux increases by $\sim 3.5$ dex, $R_{21}(I)$ and $R_{21}(T)$ increase from $\sim 0.5$ to $\sim 1.0$.
This increase is much larger than the prescription by \citet{Nara14}.
The larger scatter of $R_{21}(I)$ compared to $R_{21}(T)$ is likely due to the large scatter of $R_{21}(\sigma)$.

The positive correlation seen in top and bottom panels of Figure \ref{fig:R21_SFR} is consistent with previous studies and suggests that the ratios are elevated where molecular gas is denser and/or warmer in a close relationship to recent star formation.
From our plots and similar plots by \citet{Koda20,Yaji21}, we expect a linear relationship between the ratios and log(SFR), 
\edit1{
i.e., $R_{21} = a \times \log{\rm (SFR)} + b$.
As SFR is linearly proportional to extinction-corrected H$\alpha$ flux, we derive the slope parameter ($a = 0.152 \pm 0.005$) by fitting a line to the top panel in Figure \ref{fig:R21_SFR}.
The fitted line is shown as the orange solid line with the label ``linear'' in this figure and its slope appears to be consistent with Figure 3 of \citet{Koda20} and Figure 8 of \citet{Yaji21}.
}
Meanwhile, \citet{Ler21b} described this relationship as a power-law, \edit1{i.e., $\log{(R_{21})} = a' \times \log{\rm (SFR)} + b'$} with an index of $a' \simeq 0.13$, where both $R_{21}$ and SFR are normalized by their galactic averages.
\citet{Mae22} derived similar indices around $\simeq 0.1$ at a 100 pc resolution, with a possible variation among the galactic structures within NGC 1300.
\edit1{
We also derive the index to be $a' = 0.084 \pm 0.003$ by fitting this logarithmic function to the top panel in Figure \ref{fig:R21_SFR}.
The fitted curve is shown as the green dashed line with the label ``log'' in this figure.
Probably due to the small dynamic range in $R_{21}(I)$, the fitted two functions are very close to each other, and thus we cannot conclude which function is appropriate to describe this correlation.
Further investigation is needed to determine its formulation and to understand physical meanings of the parameters.
}

Besides the main positive correlation, we find outliers in three different regions. 
The first one is $R_{21}(I) > 1$ with moderate H$\alpha$ brightness.
As already mentioned, these data have large error bars on $R_{21}(I)$ and thus we defer further analysis to a future paper.
\edit1{
The second one is a flattening of $R_{21}(T) \simeq 0.5$ at the faint end of H$\alpha$ brightness ($\log{({\rm H}\alpha)} \lesssim 5.5$).
From the H$\alpha$ map in Figure \ref{fig:show_images}, this brightness corresponds to the peripheries of bar and spiral arms.
These components appear spatially extended and thus a contribution from a diffuse ionized gas (DIG) component is likely more important.
This flattening at the faint end suggests a possibility that DIGs and HII regions have different impact on associated molecular gas conditions.
Another possibility is that this flattening just reflects the lower limit of the ratio in the optically-thick LTE conditions.
Considering the median uncertainty in $R_{21}(T)$ of 0.14, we cannot conclude whether molecular gas is in the LTE condition or not from the current datasets.
To better estimate properties of emission lines, as discussed in \S \ref{sec:discussion}, spectral fitting is essential, especially when lines are weak.
}
The third one is the nucleus.
While H$\alpha$ is the brightest \edit1{($\log{({\rm H}\alpha)} = 9.2$)}, $R_{21}(I)$ is just above unity and $R_{21}(T) \simeq 0.85$, both of which are well below the extrapolation of the main correlation \edit1{(although it is unclear if a simple extrapolation is appropriate when ratios are close to unity, which is the upper limit in the optically-thick LTE conditions)}.
It is likely that the AGN contribution to the H$\alpha$ flux is significant \citep[e.g.,][]{Gao21} and that the AGN affects surrounding molecular gas in a different way.
However, such a significant contribution likely occurs only in this position (i.e., within a single pixel with 200 pc size), as other data points in the bright end do not remarkably deviate from the main correlation.

\edit1{
In previous studies on $R_{21}$, another type of outliers, high $R_{21}$ but with little star formation, has been reported.
As mentioned in \S \ref{sec:intro}, \citet{Sor01}  found $R_{21} \simeq 1$ in the southern part of the 30 Dor cloud complex, where star formation is not active yet.
\citet{Koda12b} mapped the entire disk of M51 at a 800 pc resolution and identified several locations in the upstream side of spiral arms, where star formation is less active compared to that of the down stream side, with $R_{21} > 0.8$.
These studies interpreted that such components are dense (but likely cold) gas before star formation.
In NGC 1365, such outliers are not clearly identified, partly due to larger errors around $R_{21}(I) \simeq 1$ and $\log(H\alpha) \lesssim 6$.
If we restrict the S/N of $R_{21}(I)$ to be $> 10$, only a few points with $R_{21}(I) > 0.75$ and $\log(H\alpha) < 5.5$ are identified. 
They are located in the bar and might represent cold and dense molecular gas with star formation suppressed due to the bar environment.
However, the current number of detection is too small to discuss the effect of bar on molecular gas conditions. 
}

On the other hand, the dependence of $R_{21}(\sigma)$ on H$\alpha$ brightness is small.
This is consistent with our finding in Figure \ref{fig:plot_ratios} that $R_{21}(\sigma) \simeq 1$ and thus $R_{21}(I) \simeq R_{21}(T)$ for most of the pixels.
\edit1{
This result indicates that 
at least at 200 pc scale, the $R_{21}(\sigma)$ scatter cannot be explained by local star formation activities.
}

\section{Discussion}\label{sec:discussion}
\edit1{
Here we focus on spectral shapes and compare line properties with moment values.
We manually select two positions with low ($R_{21}(I) \simeq 0.4$), moderate ($R_{21}(I) \simeq 0.8$), and high ($R_{21}(I) \simeq 1.2$) ratios.
}

\begin{figure*}[ht!]
\plotone{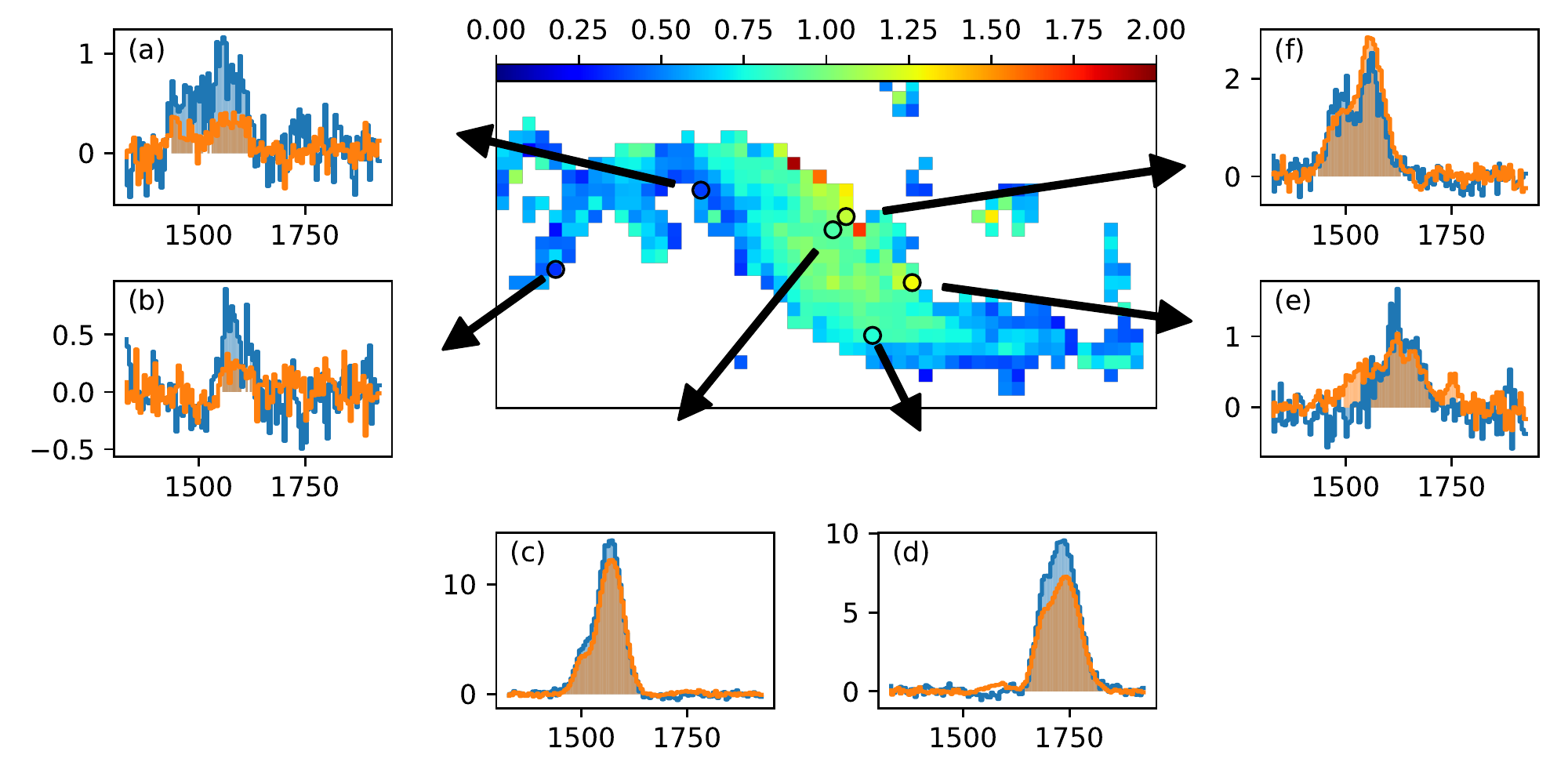}
\caption{The central panel shows a zoom-in view of $R_{21}(I)$. 
Subpanels on left present spectra in K for 
\edit1{the selected positions with $R_{21}(I) \simeq 0.4$.}
Blue and orange profiles are for CO(1--0) and CO(2--1), respectively. 
Channels used for moment calculation (i.e., \edit1{included in the final cube mask}) are shaded. 
The selected positions are indicated by open circles on the $R_{21}(I)$ map and connected to the subpanels by arrows. 
Subpanels on bottom and right are the same as those on left, but for 
\edit1{the positions with $R_{21}(I) \simeq 0.8$ and $R_{21}(I) \simeq 1.2$,} respectively.}
\label{fig:show_spectra_R21}
\end{figure*}

\begin{figure*}[ht!]
\plotone{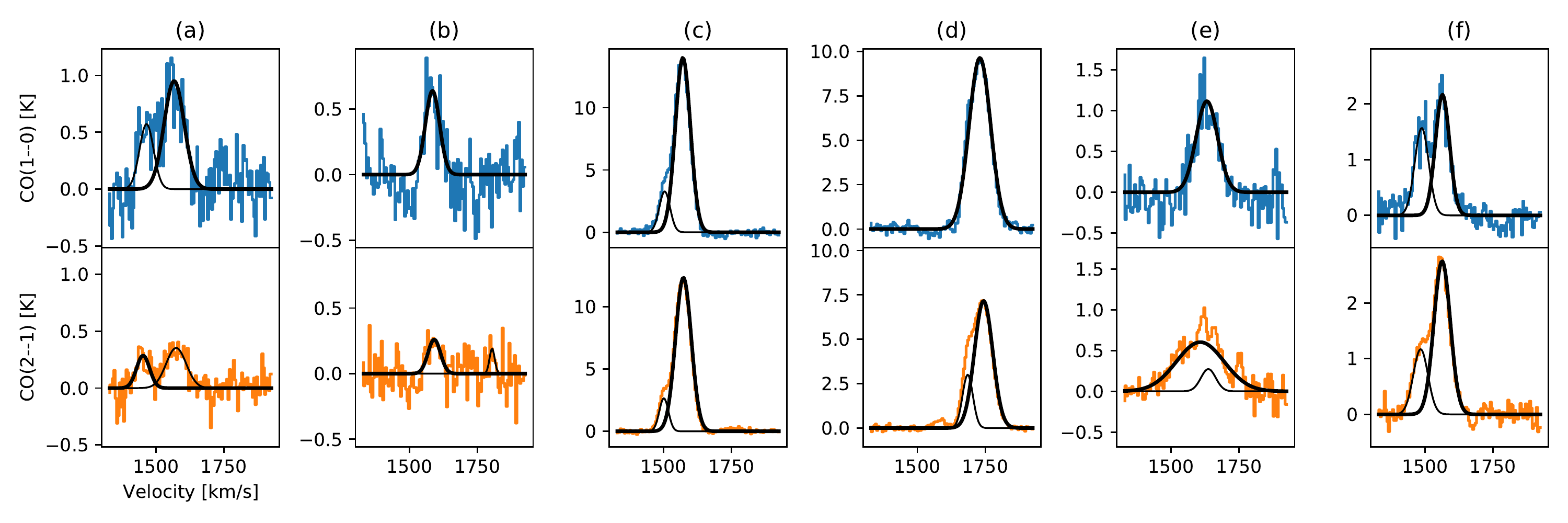}
\caption{CO(1--0) and CO(2--1) spectra of the selected positions (a)--(f) (same as shown in Figure \ref{fig:show_spectra_R21}) together with fitted Gaussian profiles (black curves).}
\label{fig:comp_fit_mom}
\end{figure*}

Figure \ref{fig:show_spectra_R21} presents the selected positions on the $R_{21}(I)$ map and their spectra.
The spectra are double peaked or show a second component in most cases, likely due to the complex dynamics explored by \citet{Gao21} and others.
Such multiple components may be in different physical conditions, which can result in different line ratios.
\edit1{To derive emission line properties more accurately under such complex environments,} we perform spectral fitting with two Gaussian components using CASA {\tt specfit}.
Fitted Gaussians are presented as solid black curves in Figure \ref{fig:comp_fit_mom}.
Fitting with two Gaussians failed for positions \edit1{(b),} (d), and (e) of CO(1--0) spectra, therefore fitting with single Gaussian was performed.
Table \ref{tab:spectra} lists fitting results for each position for each line together with moment values.
\edit1{
From this table, we find that $I$ values and the sum of fitted integrated intensities ($\Sigma I_{\rm fit}$) are consistent in most cases.
An exception happens only when the emission line is weak ($I \lesssim 50$ [K km/s]).
}
Difference between $\sigma$ and fitted line width \edit1{($V_\sigma$)} is rather complicated.
\edit1{In most cases,} $\sigma$ becomes larger than \edit1{$V_\sigma$} of each component (e.g., CO(1--0) (f)).
\edit1{
This is likely because most of the spectra comprise of multiple components.
On the other hand, $\sigma$ for the position (e) (and perhaps (b) as well) is smaller than $V_\sigma$.
Gaussian fitting might be able to better capture broad spectral features especially when the emission is weak.
}
\edit1{Nevertheless,} fitted velocity widths are approximately same between CO(1--0) and CO(2--1) except the position (e).
This is consistent with our finding based on the moment analysis, i.e.\ $R_{21}(\sigma) \simeq 1$, and thus both lines trace the same gas components in most cases.
Based on these results, if dynamics is simple (i.e., a spectrum has a single component), we claim that $R_{21}(T)$ would be the best indicator of the line ratio especially for faint emission lines.

\begin{deluxetable*}{ccDDDDDDDDRRR}
\tablecaption{Properties of selected spectra \label{tab:spectra}}
\tablehead{
\colhead{Position} & \colhead{Line} & 
\multicolumn{17}{c}{Fitting Results} & 
\colhead{$I$} & \colhead{$\sigma$} \\
\cline{3-19}
& &
\twocolhead{$V_{\rm peak}$} & \twocolhead{$V_\sigma$} &
\twocolhead{$T_{\rm peak}$} & \twocolhead{$I_{\rm fit}$} &
\twocolhead{$V_{\rm peak}$} & \twocolhead{$V_\sigma$} &
\twocolhead{$T_{\rm peak}$} & \twocolhead{$I_{\rm fit}$} & \colhead{$\Sigma I_{\rm fit}$} \\
\colhead{} & \colhead{} & 
\twocolhead{[km/s]} & \twocolhead{[km/s]} & \twocolhead{[K]} & \twocolhead{[K km/s]} &
\twocolhead{[km/s]} & \twocolhead{[km/s]} & \twocolhead{[K]} & \twocolhead{[K km/s]} &
\colhead{[K km/s]} & \colhead{[km/s]} & \colhead{[K km/s]}
}
\decimalcolnumbers
\startdata
(a) & CO(1--0) & 1567 & 37 &  1.0 &   88 & 1466 & 26 &  0.6 &   37  &  125 &  112 \pm  5 & 52 \\
(a) & CO(2--1) & 1575 & 37 &  0.4 &   33 & 1453 & 24 &  0.3 &   17  &   50 &   42 \pm  4 & 58 \\
(b) & CO(1--0) & 1584 & 26 &  0.6 &   42 & \nodata & \nodata & \nodata & \nodata  &   42 &   33 \pm  3 & 19 \\
(b) & CO(2--1) & 1590 & 23 &  0.3 &   15 & 1803 &  9 &  0.2 &    5  &   20 &   11 \pm  2 & 19 \\
(c) & CO(1--0) & 1571 & 28 & 14.0 &  978 & 1504 & 20 &  3.3 &  163  & 1141 & 1145 \pm  5 & 36 \\
(c) & CO(2--1) & 1573 & 29 & 12.3 &  901 & 1501 & 17 &  2.7 &  112  & 1013 & 1012 \pm  3 & 37 \\
(d) & CO(1--0) & 1730 & 40 &  9.6 &  965 & \nodata & \nodata & \nodata & \nodata  &  965 &  946 \pm  6 & 37 \\
(d) & CO(2--1) & 1744 & 34 &  7.2 &  603 & 1686 & 19 &  3.0 &  145  &  748 &  746 \pm  3 & 39 \\
(e) & CO(1--0) & 1632 & 40 &  1.1 &  111 & \nodata & \nodata & \nodata & \nodata  &  111 &  103 \pm  9 & 24 \\
(e) & CO(2--1) & 1607 & 88 &  0.6 &  132 & 1637 & 27 &  0.3 &   19  &  151 &  132 \pm  5 & 65 \\
(f) & CO(1--0) & 1565 & 26 &  2.2 &  141 & 1488 & 26 &  1.6 &  101  &  242 &  244 \pm  6 & 46 \\
(f) & CO(2--1) & 1563 & 30 &  2.8 &  208 & 1484 & 27 &  1.2 &   79  &  287 &  284 \pm  6 & 45 \\
\enddata
\tablecomments{(1) position label. (2) line name. 
(3)--(6) fitted peak velocity, velocity dispersion, peak temperature, and integrated intensity for the first Gaussian component. 
(7)--(10) same as (3)--(6) but for the second component. 
(11) sum of the fitted integrated intensities. (12)--(13) integrated intensity and velocity dispersion from moment calculation 
with the mask described in \S \ref{sec:mask}.
}
\end{deluxetable*}

\begin{deluxetable}{cRRRR}
\tablecaption{Measured ratios for the selected positions \label{tab:ratios}}
\tablehead{
\colhead{Position} & \colhead{$R_{21}(I)$} & 
\multicolumn{3}{c}{$R_{21}(I)_{\rm fit}$} \\
\cline{3-5}
& &
\colhead{sum} & \colhead{comp1} & \colhead{comp2}
}
\decimalcolnumbers
\startdata
(a) & 0.374 \pm 0.036 & 0.399 & 0.371 & 0.463 \\
(b) & 0.350 \pm 0.080 & 0.465 & 0.357 &   \nodata \\
(c) & 0.884 \pm 0.005 & 0.887 & 0.921 & 0.685 \\
(d) & 0.789 \pm 0.006 & 0.775 & 0.625 &   \nodata \\
(e) & 1.286 \pm 0.122 & 1.361 & 1.193 &   \nodata \\
(f) & 1.165 \pm 0.035 & 1.185 & 1.472 & 0.784 \\
\enddata
\tablecomments{(1) position label. (2) integrated line ratio from moment 0 values and its uncertainty. (3) integrated line ratio from the sum of $I_{\rm fit}$. (4) integrated line ratio for the first component in Table \ref{tab:spectra}. (5) same as (4) but for the second component if available.
}
\end{deluxetable}

\edit1{
Integrated intensity ratios calculated using the fitting results ($R_{21}(I)_{\rm fit}$ hereafter) are summarized in Table \ref{tab:ratios} together with those by moment values (i.e., $R_{21}(I)$).
}
We find that when the fit with two Gaussian components is successful (positions (a), (c), and (f)), 
\edit1{
$R_{21}(I)_{\rm fit}$ values differ between the two components.
The components with lower $R_{21}(I)_{\rm fit}$ values are with larger $V_{\rm peak}$ in (a) and with smaller $V_{\rm peak}$ in (c) and (f).
Measured velocity offsets between the two components are in the range of 70--100 km/s, which are consistent with the velocity residual from a pure circular rotation presented by \citet{Gao21}.
They proposed that this velocity residual corresponds to the outflowing gas inside or on the surface of the disk. 
According to their model (see their Figure 8), the larger $V_{\rm peak}$ component in (a) and smaller $V_{\rm peak}$ components in (c) and (f) correspond to this outflow.
}
The lower ratios suggest that density and/or temperature are lower in such outflowing gas than those in the main disk.
\edit1{
However, we should note that $R_{21}(I)_{\rm fit}$ values vary from 0.4--0.8, suggesting that physical conditions in the outflow are not uniform.
Nevertheless, this
}
difference between the two components highlights the importance of spectral decomposition for accurately measuring the integrated intensity and their ratio in particular when dynamics is complicated.

\edit1{
Meanwhile, the number of fitted components are different between the two CO lines in positions (b), (d), and (e) -- the second component is fitted only for CO(2--1) spectra.
From fitting results for (b) and (e) shown in Figure \ref{fig:comp_fit_mom}, we deduce that CO(2--1) spectra should be fitted with a single component.
Consistently, $R_{21}(I)_{\rm fit}$ values of the first component only are closer to $R_{21}(I)$ than those of sum (i.e., in Table \ref{tab:ratios}, difference between column (4) and column (2) is smaller than that between column (3) and column (2)).
The situation for (d) is rather complicated, as the two components in CO(2--1) spectra are close in velocity and one component is much brighter than the other.
While the observed spectral shapes appear similar for both lines, it remains unclear if there is really no second component in the CO(1--0) spectra.
}

To obtain a thorough view of the line ratios including their relationship with the gas dynamics, we need to perform spectral fitting in all the pixels.
\edit1{
However, the current fitting scheme is not always successful, in particular when the lines are weak.
}
A higher sensitivity will be necessary to discuss line ratios around the bar and in inter-arm regions, while a wider FoV will be necessary to investigate the ratios in the outer disk.
Furthermore, quantitatively constraining physical conditions of molecular gas requires data of other lines (e.g.\ $^{13}$CO(1--0) as done by \citet{Yaji21}).
We defer these topics to forthcoming papers.

\section{Summary}\label{sec:summary}
As the ALMA observing efficiency for CO(2--1) emission is better than that for CO(1--0), the former is now popular as a tracer of molecular gas instead of the latter.
While their ratio has often been assumed to be constant within and among galaxies, it is naturally expected to be dependent on physical conditions.
Here, we report its variation at a $2'' \simeq 200$ pc resolution in the central $2' \times 3'$ region of the nearby barred spiral galaxy NGC 1365 using ALMA data.
This FoV includes the galactic center, bar, and transition to the spiral arms.

\edit1{
A 3D cube mask is created following \citet{Mae20b} and used to make moment maps for both CO(1--0) and CO(2--1).
For calculating 2--1/1--0 ratios, another 2D mask is created based on S/N of moment 0 values and then applied.
The median value of $R_{21}(I)$, the ratio of integrated intensities, 
}
within the FoV is \edit1{0.67,} 
while the scatter is large \edit1{(0.15).} 
We also calculate the ratio of velocity dispersions (moment 2) and peak temperatures ($R_{21}(\sigma)$ and $R_{21}(T)$, respectively), and find that $R_{21}(\sigma) \simeq 1$ and thus $R_{21}(I) \simeq R_{21}(T)$ in most cases.
This result indicates that both CO(1--0) and CO(2--1) lines generally trace similar components of molecular gas. 

We create a map of extinction-corrected H$\alpha$ emission from the data provided by the PHANGS-MUSE project \citep{Ems21}.
The $R_{21}(I)$ and $R_{21}(T)$ are found to increase with the H$\alpha$ brightness, which is consistent with recent studies of nearby galaxies but at kpc-scale resolutions \citep{Koda20,Yaji21,Ler21b}.
We thus conclude that even at 200 pc resolution, $R_{21}(I)$ and $R_{21}(T)$ are elevated where the gas is denser and/or warmer due to recent star formation.
Meanwhile, high $R_{21}(I)$ (or $R_{21}(T)$) values with low H$\alpha$ brightness, which are signs of dense but cold molecular gas before star formation, are not \edit1{clearly identified.} 

\edit1{We select typical positions with low, moderate, and high ratios, and examine their spectra.}
Although the number of positions is small, we find that spectra are often double peaked and that $I$ values (i.e.\ moment 0) \edit1{slightly} underestimate true integrated intensities when the emissions are weak.
Fitting with two Gaussian components works well in most cases and its result supports our conclusion based on the moment calculations.
In addition, we identify the second component in spectra of three positions, 
and find that its integrated intensity ratio is smaller than that of the first component.
According to the kinematic model proposed by \citet{Gao21}, these components likely correspond to outflowing gas inside or on the surface of the disk.
The smaller ratios suggest that density and/or temperature of this outflowing gas are lower than those in the disk.

\acknowledgments
We are deeply grateful to an anonymous reviewer whose comments significantly improved our manuscript.
We thank Yuri Nishimura and Kotaro Kohno for a fruitful discussion.
FE, KMM, and FM are supported by JSPS KAKENHI Grant Number JP17K14259, JP19J40004, and JP21J00108, respectively.
Y.L.G gratefully acknowledges support from the China Scholarship Council (No.\ 201906340095).
GL and YG acknowledge the research grants from the China Manned Space Project (No.\ CMS-CSST-2021-A06 and No.\ CMS-CSST-2021-A07), the National Natural Science Foundation of China (No.\ 11421303), the Fundamental Research Funds for the Central Universities (No.\ WK3440000005), and the K.\ C.\ Wong Education Foundation.
This paper makes use of the following ALMA data: ADS/JAO.ALMA\#2015.1.01135.S, ADS/JAO.ALMA\#2017.1.00129.S, ADS/JAO.ALMA\#2013.1.01161.S. ALMA is a partnership of ESO (representing its member states), NSF (USA) and NINS (Japan), together with NRC (Canada), MOST and ASIAA (Taiwan), and KASI (Republic of Korea), in cooperation with the Republic of Chile. 
The Joint ALMA Observatory is operated by ESO, AUI/NRAO, and NAOJ.
Data reduction was carried out on the Multi-wavelength Data Analysis System operated by the Astronomy Data Center (ADC), National Astronomical Observatory of Japan.
We would like to thank Editage (www.editage.com) for English language editing.

\vspace{5mm}
\facilities{ALMA, VLT/MUSE}

%% just to avoid appendix overlapping acknowledgments
\clearpage

%% references %%
%\bibliography{papers}{}
%\bibliographystyle{aasjournal}

%% for arxiv: copied from output.bbl

\end{document}